\documentclass[print]{aa}
\usepackage{graphicx}

\usepackage[varg]{txfonts}
\usepackage{cases}
\usepackage{natbib}
\usepackage{underscore}
\usepackage{hyperref}

\usepackage{amsmath}

\defcitealias{PerriLeitner2022}{Perri \& Leitner et al.} 
\defcitealias{Mikic1999}{Miki\'c et al., 1999} 
\defcitealias{Kuzma2023}{Ku\'zma et al., 2023} 

\begin{document}
      \title{Modelling the propagation of coronal mass ejections with COCONUT: implementation of the Regularized Biot-Savart Laws flux rope model}
      
      \author{J. H. Guo\inst{1,2}, L. Linan\inst{2}, S. Poedts\inst{2,3}, Y. Guo\inst{1}, A. Lani\inst{2}, B. Schmieder\inst{2,4}, M. Brchnelova\inst{2}, B. Perri\inst{5,2}, T. Baratashvili\inst{2}, Y. W. Ni\inst{1}, P. F. Chen\inst{1}}
      
      \institute{School of Astronomy and Space Science and Key Laboratory of Modern Astronomy and Astrophysics, Nanjing University, Nanjing 210023, China \\
      	\email{chenpf@nju.edu.cn}
 	    \and 
 	    Centre for Mathematical Plasma Astrophysics, Department of Mathematics, KU Leuven, Celestijnenlaan 200B, B-3001 Leuven, Belgium \\
        \email{Stefaan.Poedts@kuleuven.be}
        \and
        Institute of Physics, University of Maria Curie-Skłodowska, ul.\ Radziszewskiego 10, 20-031 Lublin, Poland
        \and
        LESIA, Observatoire de Paris, CNRS, UPMC, Universit\'{e} Paris Diderot, 5 place Jules Janssen, 92190 Meudon, France
        \and
        AIM/DAp - CEA Paris-Saclay, Université Paris-Saclay, Université Paris-Cité, Gif-sur-Yvette, France
 	    }
    \titlerunning{RBSL flux rope model in COCONUT}
 \authorrunning{Guo et al.}
\date{}

\abstract
{Coronal mass ejections (CMEs) are rapid eruptions of magnetized plasma that occur on the Sun, which are known as the main drivers of adverse space weather. Accurately tracking their evolution in the heliosphere in numerical models is of utmost importance for space weather forecasting.}
{The main objective of this paper is to implement the Regularized Biot-Savart Laws (RBSL) method in a new global corona model COCONUT. This approach has the capability to construct the magnetic flux rope with an axis of arbitrary shape.} 
{We present the implementation process of the RBSL flux rope model in COCONUT, which is superposed onto a realistic solar wind reconstructed from the observed magnetogram around the minimum of solar activity. Based on this, we simulate the propagation of an S-shaped flux rope from the solar surface to a distance of 25$R_{\odot}$.}
{Our simulation successfully reproduces the birth process of a CME originating from a sigmoid in a self-consistent way. The model effectively captures various physical processes and retrieves the prominent features of the CMEs in observations. In addition, the simulation results indicate that the magnetic topology of the CME flux rope at around 20$R_{\odot}$ deviates from a coherent structure, and manifests as a mix of open and closed field lines with diverse footpoints.}
{This work demonstrates the potential of the RBSL flux rope model in reproducing CME events that are more consistent with observations. Moreover, our findings strongly suggest that magnetic reconnection during the CME propagation plays a critical role in destroying the coherent characteristic of a CME flux rope.}

\maketitle
\nolinenumbers

\section{Introduction}\label{introduction} 

Coronal mass ejections (CMEs) are the most remarkable eruptions observed within the solar system. These powerful ejections can expel substantial quantities of magnetized plasma from the Sun to the interplanetary space, profoundly impacting the heliospheric environment \citep{Chen2011, Webb2012, Schmieder2015}. In general, they serve as the primary drivers of adverse space weather, such as geomagnetic storms and gradual solar energetic particle (SEP) events. These could hinder satellite operations and pose risks to human health \citep{Gosling1993, Schrijver2015}. Furthermore, its early evolution in the solar corona and the subsequent propagation through the interplanetary space involve abundant physical processes, e.g., magnetic reconnection, heating, plasma waves, and particle acceleration \citep{Tsurutani2023}. The significance of these processes extends beyond CMEs and to a wide range of eruptive phenomena observed across various celestial bodies, from (exo)-planet magnetospheres to black holes. Taken as a whole, studies of CMEs not only enable the improvement of forecasting capabilities regarding detrimental space weather events, but also enhance our understanding on the fundamental astrophysical and plasma processes.

Albeit CMEs have been observed for several decades, their involved scientific issues and accurate predictions have not been addressed in a very satisfactory way yet due to limited observations \citep{Chen2011}. On the one hand, CMEs are generally observed above the solar limb by the white-light coronagraphs occulting the low corona \citep{Illing1986}, indicating that their initial processes and evolution in the early stage are hard to be tracked. On the other hand, the radiation mechanism of the white light poses difficulty to derive the temperature of CMEs \citep{Vourlidas2006}. Moreover, the magnetic fields of CMEs cannot be directly measured yet, despite the fact that it is the basis to predict the geomagnetic effects of their consequent interplanetary counterparts, especially the orientation and intensity of their southward component for their interaction with Earth's magnetosphere. To address these limitations, significant advances have been made in numerical MHD modelling in recent years, such as the models of ENLIL \citep{Odstrcil2003}, SWMF framework \citep{Toth2012}, EUHFORIA \citep{Pomoell2019, Poedts2020}, SIP-CESE \citep{Feng2007, Zhou2012, Feng2020}, MS-FLUKSS \citep{Singh2018}, AwSoM \citep{VanderHolst2010, Jin2017}, ICARUS \citep{Verbeke2022, Baratashvili2022}, SUSANOO \citep{Shiota2016}, COIN-TVD \citep{Shen2014}, and MAS \citep{Mikic2018}. These numerical tools play an indispensable role in both space-weather forecasting goals and scientific understandings. 

Many observations and theories suggest that magnetic flux ropes play a pivotal role in explaining the generation and subsequent evolution of CMEs. For example, the statistical studies conducted by \citet{Ouyang2017} indicated that about $\sim$89\% of erupting filaments are supported by flux ropes, meaning that flux ropes already exist prior to eruptions in a majority of CMEs. Besides, \citet{Vourlidas2013} suggested that at least 40\% of CMEs observed by white-light coronagraphs exhibit typical flux-rope structures. When CMEs arrive at the interplanetary space and are detected by in-situ satellites, their carrying magnetic fields often exhibit large and smooth rotations, which are generally considered as the proxy of a helical flux rope \citep{Burlaga1981}. In addition, preexisting flux ropes are often indispensable in many models to explain the CME initiation (albeit not always), such as the kink instability \citep{Hood&Priest1981, Torok2004} and torus instability \citep{Kliem2006, Aulanier2010}. Therefore, several forecasting models place emphasis on the construction of flux ropes to initiate CMEs. 

Nowadays, the extensively used flux-rope models to trigger CMEs can be mainly divided into two categories. The first type is the sphere-like models without legs, where the representative ones are the spheromak model \citep{Kataoka2006, Shiota2010, Verbeke2019} and the tear-drop Gibson-Low model \citep{Gibson1998}. The second category is the toroidal flux rope anchored to the solar surface, such as the ``Flux Rope in 3D" \citep[FRi3D;][]{Isavnin2016}, Titov-D{\'e}moulin \citep[TD;][]{Titov1999}, and Titov-D{\'e}moulin-modified models \citep[TDm;][]{Titov2014}, which should be more effective in reproducing flank-encounter events \citep{Maharana2021}. 

Even so, the aforementioned flux rope models still deviate considerably from the observed flux ropes, which usually exhibit sigmoids, U-shaped, or more irregular morphology \citep{Cheng2017}. It is thus essential to develop a method to construct a flux rope whose morphology aligns more accurately with the observations. For this purpose, \citet{Titov2018} proposed an advanced method to construct a force-free flux rope in the source region, called Regularized Biot-Savart Laws (RBSL) method. This enables the construction of a flux rope with an axis of arbitrary shape, making it possible to accurately reproduce the CMEs originating from the flux ropes with irregular shapes. Based on this method, we successfully reconstructed the magnetic structures of some filaments \citep{Guoy2019, Guojh2021}, and model the onset processes of their resulting CMEs in an observational data-driven way \citep{Guoy2021, Guojh2023, Guoy2023}. As such, it is expected that the RBSL flux rope model has the potential in improving the accuracy of space weather forecasting.

On the other hand, in many heliospheric space-weather forecasting simulations, the CMEs are generally introduced beyond the super-Alfv\'enic point, typically at around 0.1 AU \citep{Verbeke2019, Scolini2019, Scolini2020, Maharana2021, Maharana2023}. The input parameters of CMEs are determined based on the coronagraph observations, with the assumption that CMEs evolve self-similarly in the corona up to 0.1~AU. Although such an assumption has achieved significant success in contemporary space weather predictions, it is noticed that CMEs often undergo rotation and deflection in this region \citep{Shen2022}. For example, \citet{Lynch2009} found that the rotation angle can reach 50$^\circ$ at 3.5$R_{\odot}$. \citet{Shiota2010} found that the reconnection between the eruptive flux rope and ambient magnetic field lines can lead to the rotation of the CME flux rope within 5$R_{\odot}$. The observational data-constrained simulation performed by \citet{Guojh2023b} suggested that magnetic reconnection during the eruption can result in reconstruction of the flux rope axis, manifested as the lateral drifting of filament materials in observations. The heliospheric MHD simulations from the solar surface to 1 AU revealed that the rotation of the flux rope is likely to occur within 15$R_{\odot}$ \citep{Regnault2023}. Observations also demonstrated that the magnetic-field gradient can result in the orientation change of the flux rope axis \citep{Shen2011, Gui2011, Liuy2018}, and thus induce the consequent otherwise unexpected strong geomagnetic storms. \citet{Asvestari2022} found that the tilt angles between the flux-rope axes and their surrounding background fields can affect their rotation angle when propagating through the heliosphere. Therefore, to input a more realistic CME for the interplanetary simulation, it is essential to track the evolution of the flux rope within 25$R_{\odot}$. As demonstrated in some previous studies \citep{Jin2012, Jin2017, Zhou2017, Torok2018}, where the CMEs modeled in the corona module are integrated into the heliosphere simulations, the corona-interplanetary space coupling models exhibit the potential to improve the accuracy of the space weather prediction.

The main objective of this paper is to implement the RBSL flux rope model into our recently developed three-dimensional (3D) global coronal model, called COolfluid COroNal UnsTructured (COCONUT, \citetalias{PerriLeitner2022}~\citeyear{PerriLeitner2022}), and then simulate its propagation process from the solar surface to a distance of 25$R_{\odot}$. In our former work \citep{Linan2023}, we simulated the CME initiated from a toroid-shaped TDm flux rope. Built upon that foundation, this work will employ the RBSL flux ropes to drive CMEs, enabling us to model the events erupting from the flux ropes characterized by more realistic shapes, such as the sigmoid. This paper is organized as follows. The modelling description is introduced in Section \ref{sec:met}, and the results are exhibited in Section \ref{sec:res}, which are followed by discussions and a summary in Sections \ref{sec:dis} and \ref{sec:sum}, respectively.

\section{Modelling description}\label{sec:met}

\subsection{Infrastructure of the COCONUT solver}

In this paper, we utilize COCONUT, a state-of-the-art MHD solver for global solar coronal modelling recently developed by \citetalias{PerriLeitner2022}~(\citeyear{PerriLeitner2022}), to model the propagation of CMEs. This code is built upon the Computational Object-Oriented Libraries for Fluid Dynamics (COOLFluiD) platform \citep{Kimpe2005, Lani2005, Lani2013, Lani2014}, which enables the solution of 3D full MHD equations using an implicit scheme. The adoption of the implicit scheme allows the disregard of constraints imposed by the Courant-Friedrichs-Lewy (CFL) requirement in explicit solvers, enabling the convergence of steady-state solutions at a significantly faster speed. Moreover, the utilization of unstructured meshes eliminates the polar singularity, facilitating complete coverage of the solar corona including the solar poles \citep{Brchnelova2022}. It is noticed that, \citet{Perri2023} demonstrated the importance of considering the solar poles even for the ecliptic-plane prediction. Furthermore, the implementation of unstructured meshes in COCONUT makes it well-suited for advanced numerical techniques for grid refinement and the high-order Flux Reconstruction method \citep{Vandenhoeck2019}. These advanced approaches enhance the capabilities and potential of COCONUT in space-weather forecasting applications.

In line with \citet{Linan2023}, we employ a polytropic MHD model to simulate the CME propagation in the inner heliosphere. This model incorporates an adiabatic energy equation with a reduced adiabatic index, as previously described by \citet{Mikic1999} and \citetalias{PerriLeitner2022}~(\citeyear{PerriLeitner2022}). The governing equations are as follows:

\begin{eqnarray}
 && \frac{\partial \rho}{\partial t} +\nabla \cdot(\rho \boldsymbol{v})=0,\label{eq1}\\
 && \frac{\partial (\rho \boldsymbol{v})}{\partial t}+\nabla \cdot(\rho \boldsymbol{vv}+\boldsymbol{I}(p+\frac{1}{2}|B|^2)-\boldsymbol{BB})=\rho \boldsymbol{g},\label{eq2}\\
 && \frac{\partial \boldsymbol{B}}{\partial t} + \nabla \cdot(\boldsymbol{vB-Bv}+\boldsymbol{I}\phi)=0,\label{eq3}\\
 && \frac{\partial E}{\partial t}+\nabla \cdot((E+p+\frac{1}{2}|\boldsymbol{B}|^2)\boldsymbol{v}-\boldsymbol{B(v \cdot B)})  =\rho \boldsymbol{g \cdot v}, \label{eq4}\\
 && \frac{\partial \phi}{\partial t}+\nabla \cdot(V_{\rm ref}^{2}\boldsymbol{B})=0, \label{eq5}
\end{eqnarray}
where $p$ is the thermal pressure, $E=\rho v^{2}/2+p/(\gamma -1)+B^2/8\pi$ is the total energy density, $\boldsymbol{g}(r)=-(GM_{\odot}/r^2)\hat{\boldsymbol{e_{r}}}$, is the gravitational acceleration, $\gamma=1.05$ is the reduced adiabatic index, $\phi$ is used to clean the magnetic-field divergence, and the other parameters have the usual meanings. Equation (5) is introduced by the hyperbolic divergence cleaning method \citep{Dedner2002}, which is employed to mitigate the divergence of magnetic fields arising during numerical calculations. It is noteworthy that the motivation for using this reduced $\gamma$ value is based on the fact that the coronal temperature does not change substantially, which mimics the quasi-isothermal heating with limited energy injection. As demonstrated in previous works (\citetalias{Mikic1999}, \citetalias{PerriLeitner2022}~\citeyear{PerriLeitner2022}, \citetalias{Kuzma2023}), the polytropic model can successfully drive either the slow and fast solar wind (but not the bimodal distribution), and reproduce large-scale streamers and the coronal hole distribution in observations, although it is difficult to maintain the thermal properties consistent with the solar corona, such as plasma $\beta$ \citep{Regnault2023}. Nevertheless, the evolution of magnetic topology remains realistic with the full resolution of the MHD equations.

The simulation domain covers the whole global corona including the two poles, where $r$ spans from the solar surface to 25$R_{\odot}$ in radius, $\theta$ ranges from $-90^{\circ}$ to $90^{\circ}$ in latitude, and $\varphi$ ranges from $-180^{\circ}$ to $180^{\circ}$ in longitude. To discretize this domain, we use a 6-level subdivision of the geodesic polyhedron, resulting in approximately 1.5 million cells. Further details regarding the effects of unstructured meshes in COCONUT can be found in \citet{Brchnelova2022}. For the boundary conditions, we follow the prescription outlined in \citetalias{PerriLeitner2022}~(\citeyear{PerriLeitner2022}). At the inner boundary, the radial component of the magnetic field is determined by the imposed magnetogram, while the other components are extrapolated in a zero-gradient manner. The density and pressure at the inner boundary are set to fixed values of $\rho_{\odot} = 1.67 \times 10^{-16}$ g cm$^{-3}$ and $p_{\odot} = 4.15 \times 10^{-2}$ dyn cm$^{-2}$, respectively. Additionally, we employ the technique described in \citet{Brchnelova2022b} to reduce the generation of spurious electric fields. As such, the velocity on the surface of the Sun is prescribed with a small outflow that aligns with the magnetic field lines. Regarding the outer boundary, the physical quantities are extrapolated in a zero-gradient manner. More comprehensive information and numerical tests regarding the boundary prescription can be found in \citetalias{PerriLeitner2022}~(\citeyear{PerriLeitner2022}).

Our simulation consists of two steps. Firstly, we construct the background solar wind where the coronal plasmas move outward, by utilizing the time-independent relaxation module for the steady-state solution of the solar wind developed in \citetalias{PerriLeitner2022}~(\citeyear{PerriLeitner2022}). Once this background is established, we transition to a time-dependent model to simulate the evolution of CMEs. Following the approach presented by \citet{Linan2023}, we employ a three-point backward time discretization scheme with a time step of 0.01 in the normalization unit (equivalent to 14.4 seconds in physical units). The convergence tests indicate that the time resolution primarily have slight impacts on the magnitude of plasma profiles while keeping the trend. Hence, it is considered acceptable to adopt this resolution as a benchmark to model CMEs, taking into account the trade-off between the computational speed and solution accuracy. For further insights into the specifics of this numerical scheme and additional testing, please refer to \citet{Linan2023}. In the following subsections, we will present more details about the description of the numerical setup process.

\subsection{Quasi-steady solution of the background solar wind}

In this subsection, we outline the procedure for modeling the background solar wind. We use the magnetogram of 2019 July 2, observed by the Helioseismic and Magnetic Imager \citep[HMI;][]{Scherrer2012} onboard the Solar Dynamics Observatory \citep[SDO;][]{Pesnell2012}, which can provide a more realistic solar wind for the CME propagation compared to the dipole. The reasons why we choose the data near solar minimum are as follows. Firstly, this date coincides with a total solar eclipse on Earth, providing unique observations to validate the model. In addition, the solar-wind configuration based on this magnetogram has been extensively examined in other COCONUT papers \citep{Perri2023, Kuzma2023, Linan2023}. Furthermore, the magnetic configuration of the solar minimum activity is comparatively simpler such that the convergence is easier to be obtained compared to the case of solar maximum activity. To smooth the input magnetogram and thereby enhance the numerical stability, following our previously established pretreatment procedure (\citetalias{PerriLeitner2022}~\citeyear{PerriLeitner2022}, \citetalias{Kuzma2023}), the original magnetogram is preprocessed through the projection onto spherical harmonics, with a maximum frequency for reconstruction of $l_{\rm max}=20$.

The initial magnetic fields are provided by the potential-field source-surface (PFSS) model, which is computed with a fast Finite Volume solver for the Laplace equation in COCONUT (\citetalias{PerriLeitner2022}~\citeyear{PerriLeitner2022}). The initial conditions for density and pressure are set in accordance with \citet{Brchnelova2022b}. Subsequently, we initiate a relaxation process using the polytropic MHD equations Eqs. (\ref{eq1}--\ref{eq5}). To assess convergence, we utilize the global residuals of various physical quantities, which are evaluated using the following formula:

\begin{eqnarray}
{\rm res}(a)={\rm log}\sqrt{\sum_{i}(a_{i}^{t}-a_{i}^{t+1})^{2}}, \label{eq6}
 \end{eqnarray}
where $a$ is the considered physical quantities in normalization unit, $i$ is the spatial index, and $t$ is the temporal index. Correspondingly, the mean iteration change in physical unit can be described as $\overline{da}=a_{0}10^{\rm res(a)}/\sqrt N_{\rm cell}$, where $a_{0}$ denotes the normalization unit ($l_{0}=6.955\times 10^{10}$~cm, $\rho_{0}=1.67\times10^{-16}$~g cm$^{-3}$, $B_{0}=2.2$~G, $v_{0}=4.8\times10^{7}$~cm s$^{-1}$), and $N_{\rm cell}$ represents the number of cells. After approximately 15000 iterations, a convergence level of $-4.5$ in the radial velocity is attained (res($v$)=-4.5), indicating that the average variation of radial velocity during the iteration, $\overline{dv}$, is close to 1~cm s$^{-1}$. Moreover, as demonstrated in \citet{Linan2023}, the density map of the modeled solar wind exhibits a certain similarity to that derived from the observations with the tomography method \citep{Morgan2015}. Therefore, based on our previous numerical tests (\citetalias{PerriLeitner2022}~\citeyear{PerriLeitner2022}, \citetalias{Kuzma2023}), the obtained solar wind solution can be roughly characterized as a convergence state in our model.

\begin{figure*}[htbp]
  \includegraphics[width=15cm,clip]{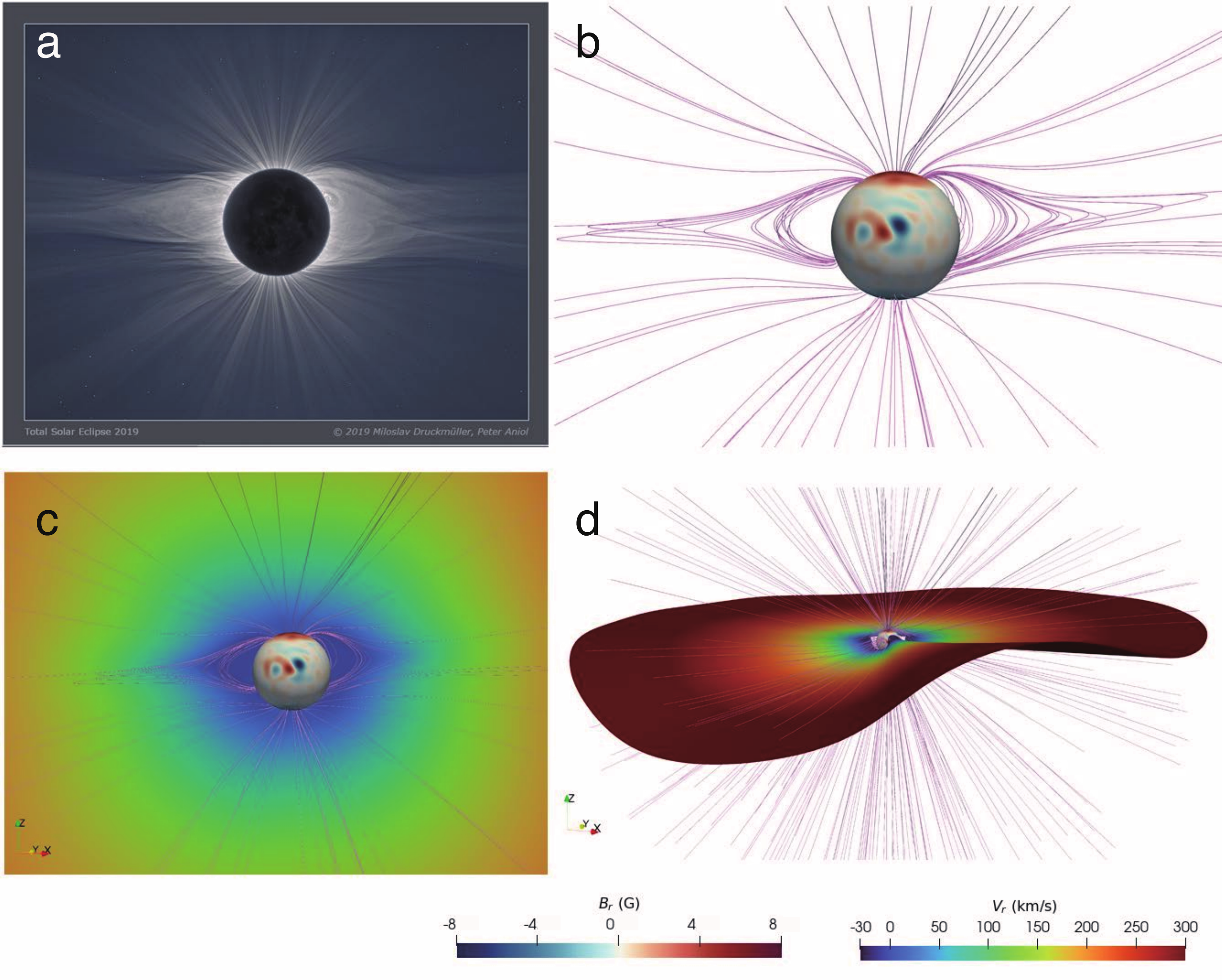}
  \centering
  \caption{Comparisons between observations and numerical results for the background solar wind. Panel~(a) displays the solar eclipse image of 2nd July 2019 (Copyright: 2019 Miloslav Druckm\"{u}ller, Peter Aniol), corresponding to a minimum of solar activity. Panels~(b) and (c) correspond to the typical field lines and radial velocity distribution in the meridian plane, respectively. Panel~(d) showcases the visualization of the heliospheric current sheet depicted by the iso-surface of $B_{r}=0$. \label{fig1}}
\end{figure*}

Figure~\ref{fig1} presents a comparison between the observations and numerical results. In Panel~(a), white-light images assembled from 128 eclipse pictures are displayed \citep{Boe2020}. The finest details of the corona, including equatorial streamers stretching radially on both sides and open streamers extending from the polar regions, are prominently featured. Panel~(b) illustrates the typical fields in our solar wind model, roughly rotated to the viewing angle from Earth on the date of the eclipse. It is seen that the streamers depicted in our COCONUT model closely match the observations in terms of morphology and placement. The radial velocity profile is showcased in Panel~(c), with two large equatorial streamers visible, in agreement with the distribution of field lines. Panel~(d) visualizes the 3D structure of the heliospheric current sheet (HCS) using an iso-surface of $B_{r}=0$. For further details regarding the validation process, please refer to the comprehensive study conducted by \citet{Kuzma2023} and \citet{Linan2023}. Overall, the agreement between the observed coronal streamers and the magnetic field structures in the simulation indicates the reasonableness of our coronal model.

\subsection{Implementation of the RBSL flux rope model}
Once the background solar wind reaches a relatively steady state, we proceed to superpose a flux rope into it to model the eruption and propagation of a CME. In this study, we employ the RBSL method proposed by \citet{Titov2018} to construct the preexisting flux rope. Unlike the TDm model which was implemented by \citet{Linan2023}, they superposed a flux rope fixed in a toroidal shape, while the RBSL method allows for the construction of a flux rope with an arbitrary path for its axis. As a result, it is well-suited for simulating CMEs originating from complex morphologies in the source region, such as sigmoids. In the following paragraphs, we will detail the implementation process of the RBSL flux rope in COCONUT.

Essentially, the RBSL flux rope is an electric current channel carrying the axial current $I$ with a parabolic profile.  Its magnetic field can be described using the following formulae:
\begin{eqnarray}
\boldsymbol{B}=\nabla \times \boldsymbol{A_{I}} + \nabla \times \boldsymbol{A_{F}}, \label{eq7} \\
\boldsymbol{A_{I}(x)}=\frac{\mu I}{4\pi}\int_{C\cup C^{'}}K_{I}(r)\boldsymbol{R^{'}}(l)\frac{dl}{a(l)}, \label{eq8} \\
\boldsymbol{A_{F}(x)}=\frac{F}{4\pi}\int_{C\cup C^{'}}K_{F}(r)\boldsymbol{R^{'}}(l)\times\boldsymbol{r}\frac{dl}{a(l)^2}, \label{eq9}
 \end{eqnarray}
where $a(l)$ represents the minor radius of the flux rope, $C$ represents its axis path, $C^{'}$ is the mirror path of the axis with respect to the photosphere, $l$ represents the arc length, $\boldsymbol{R(l)}$ is the radius-vector, $\boldsymbol{R^{'}}$ is the tangential unit vector, $\boldsymbol{r}=(\boldsymbol{x}-\boldsymbol{R(l)})/a(l)$ represents the vector from the source point to the field point. The kernels of the Regularized Biot-Savart laws, denoted as $K_I(r)$ and $K_F(r)$, are described by the piecewise functions that incorporate the internal and external solutions \citep{Titov2018}. The expressions are as follows:

\begin{eqnarray}
 K_{I}(r)=  \begin{cases} \frac{2}{\pi}(\frac{\rm{arcsin r\ }}{r}+\frac{5-2r^{2}}{3} \sqrt{1-r^{2}}) \qquad 0 \le r \le 1 \\
 \frac{1}{r} \qquad  r > 1, \end{cases} \label{eq10}
 \end{eqnarray}

\begin{eqnarray}
 K_{F}(r)=  \begin{cases} \frac{2}{\pi r^{2}}(\frac{\arcsin r}{r}-\sqrt{1-r^{2}})+\frac{2}{\pi}\sqrt{1-r^{2}}
 + \\
 \frac{5-2r^{2}}{2\sqrt{6}}[1-\frac{2}{\pi} \arcsin(\frac{1+2r^{2}}{5-2r^{2}})] \qquad  0 \le r \le 1,\\ 
 \frac{1}{r^{3}} \qquad  r > 1, \end{cases} \label{eq11}
 \end{eqnarray}
where the kernels in the region $r>1$ correspond to the classic Biot-Savart law, and those for $0 \le r\le 1$ correspond to the Regularized Biot-Savart law \citep{Titov2018}. This is derived from the force-free condition, assuming a parabolic distribution of electric current density. In this formulation, the relationship between the axial flux $F$ and the axial current $I$ satisfies $F=\pm(3 \mu_{0} Ia)/(5\sqrt{2})$, with the sign determined by the helicity of the flux rope. The positive (negative) sign corresponds to a flux rope with positive (negative) helicity, where the field lines wrap around the axis following the right-hand (left-hand) rule. 

Eqs. (\ref{eq7}--\ref{eq11}) outline the fundamental principles for constructing an RBSL flux rope, which is primarily governed by four key parameters: the flux rope path ($C$), the minor radius ($a$), the axial flux ($F$), and the electric current ($I$). Once these parameters are determined, the RBSL flux rope can be uniquely constructed. Hence, the initial step is to define the flux rope path. The main objective of this paper is to implement the RBSL flux rope in COCONUT and model the propagation of its resulting CME. For simplicity, we employ a theoretical curve to govern the morphology of the flux-rope axis, rather than using the path directly measured from observations. This means that we do not concentrate on a specific observed event in this work. The insertion of the flux rope will lead to two additional poles with strong magnetic fields in the original magnetogram. In future papers, we will reproduce a truly observed event, in which the construction of the flux rope is highly constrained by the observations and more comparisons with the observations are conducted. Following \citet{Torok2010} and \citet{Xu2020}, we adopt the following equations to control the path of the flux rope:

 \begin{eqnarray}
 && f(s)=  \begin{cases} \frac{s(2x_{c}-s)}{x_{c}^{2}}\theta, \quad 0 \le s \le x_{\rm c} \\ \frac{(s-2x_{_{\rm c}}+1)(1-s)}{(1-x_{c})^{2}}\theta, \qquad x_{_{c}} < s \le 1 \end{cases} \label{eq12}
  \end{eqnarray}
\begin{eqnarray}
   x=(s-x_{c})\cos f + x_{c}, \label{eq13}\\
   y=(s-x_{c})\sin f, \label{eq14}
 \end{eqnarray}
 
 \begin{eqnarray}
 && z(x)=  \begin{cases} \frac{x(2x_{\rm h}-x)}{x_{\rm h}^{2}}h, \quad 0 \le x \le x_{\rm h} \\ \frac{(x-2x_{_{\rm h}}+1)(1-x)}{(1-x_{\rm h})^{2}}h, \qquad x_{_{\rm h}} < x \le 1 \end{cases} \label{eq15}
 \end{eqnarray}
where $x_{c}$ controls the intersection of the projected curve and the line connecting two footpoints (defined as the crossing point herein), the angle $\theta$ controls the orientation of the tangent vector at the crossing point, $x_{h}$ controls the position of the apex, and $h$ determines the apex height. In particular, since the RBSL method requires a closed path for the current circuit, we add a mirror sub-photosphere path to close it. Once the path is determined, we need to set the minor radius $a$, axial flux $F$, and then the electric current $I$ can be derived from the force-free condition. Subsequently, we can obtain the kernels of the RBSL using Equations (\ref{eq10}) and (\ref{eq11}). Finally, the magnetic fields of the flux rope can be calculated using Equations (\ref{eq7})--(\ref{eq9}).

\begin{figure*}[htbp]
  \includegraphics[width=\textwidth,clip]{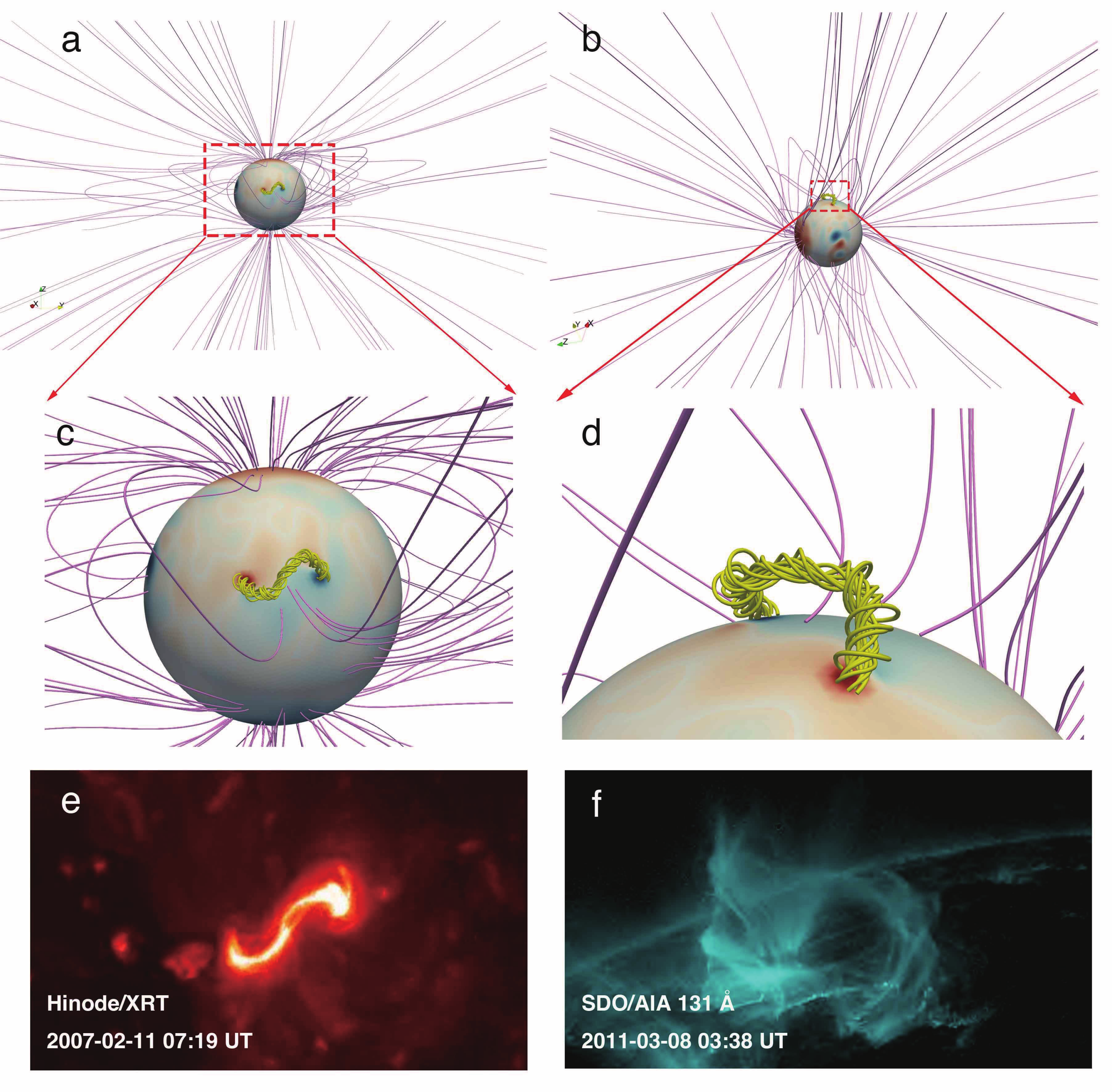}
  \centering
  \caption{Visualization of the magnetic field lines of the global corona coupled with the RBSL flux rope (yellow tubes) with the background solar wind (pink tubes). Panels~(a) and (b) illustrate the results viewed from the top and side, respectively. Panels~(c) and (d) show the zoomed-in views of the red rectangles in Panels~(a) and (b), respectively. Panels~(e) and (f) showcase the typical flux rope in observations, which are reflected by the sigmoid viewed on the solar disk and the hot channel at the solar limb. The sigmoid on the solar disk in panel~(e) is observed by the Hinode/XRT. The hot channel above the solar limb in panel (f) is observed by the SDO/AIA 131 \AA\ band.} \label{fig2}
\end{figure*}

Figure~\ref{fig2} showcases the RBSL flux rope constructed with the following parameter values: $x_{c}=x_{h}=0.5$, $\theta=60^{\circ}$, $h=120\;$Mm, $a=35\;$Mm, and $F=3\times10^{20}\;$Mx. The flux rope is inserted at the equator with a longitude of $30^{\circ}$, situated within the solar quiet region. The selected electric current of the flux rope is about 10 times the intensity estimated by Shafranov’s equilibrium equation (Equation (7) in \citeauthor{Titov2014}~\citeyear{Titov2014}), meaning that the flux rope is unstable and will undergo a direct eruption upon insertion. We do not change the thermodynamic properties within the flux rope, encompassing the velocity fields, pressure and density, such that the eruption is fully propelled by the disequilibrium of the magnetic fields. 

To validate our modeled flux rope, we perform a comparison with the flux-rope proxies in observations, such as sigmoids and hot channels \citep{Zhang2012, Cheng2013}. Figure~\ref{fig2}e presents a sigmoid on the solar disk observed by the Hard X-Ray Telescope (XRT) onboard Hinode \citep{Golub2007}. It is immediately clear that the modeled flux rope, when viewed from the top (Figure~\ref{fig2}c), closely resembles the observed sigmoid. Regarding side views, Figure~\ref{fig2}f displays a hot channel observed by the SDO/AIA 131 \AA\ band \citep{Cheng2013}. As depicted in Figures~\ref{fig2}d and \ref{fig2}f, both the modeled flux rope and the observed hot channel display a similar arch shape. Therefore, the flexibility of the RBSL flux-rope path can effectively reduce the deviations from observed progenitors of CMEs in solar source regions. Despite the flux-rope path being determined by a theoretical curve in this work, the comparability between the modeled flux rope and observations demonstrates the capability of the RBSL method in reconstructing flux ropes that are consistent with observations.

\section{Numerical Results}\label{sec:res}

\subsection{Global evolution of the CME flux rope}

\begin{figure*}
  \includegraphics[width=\textwidth,clip]{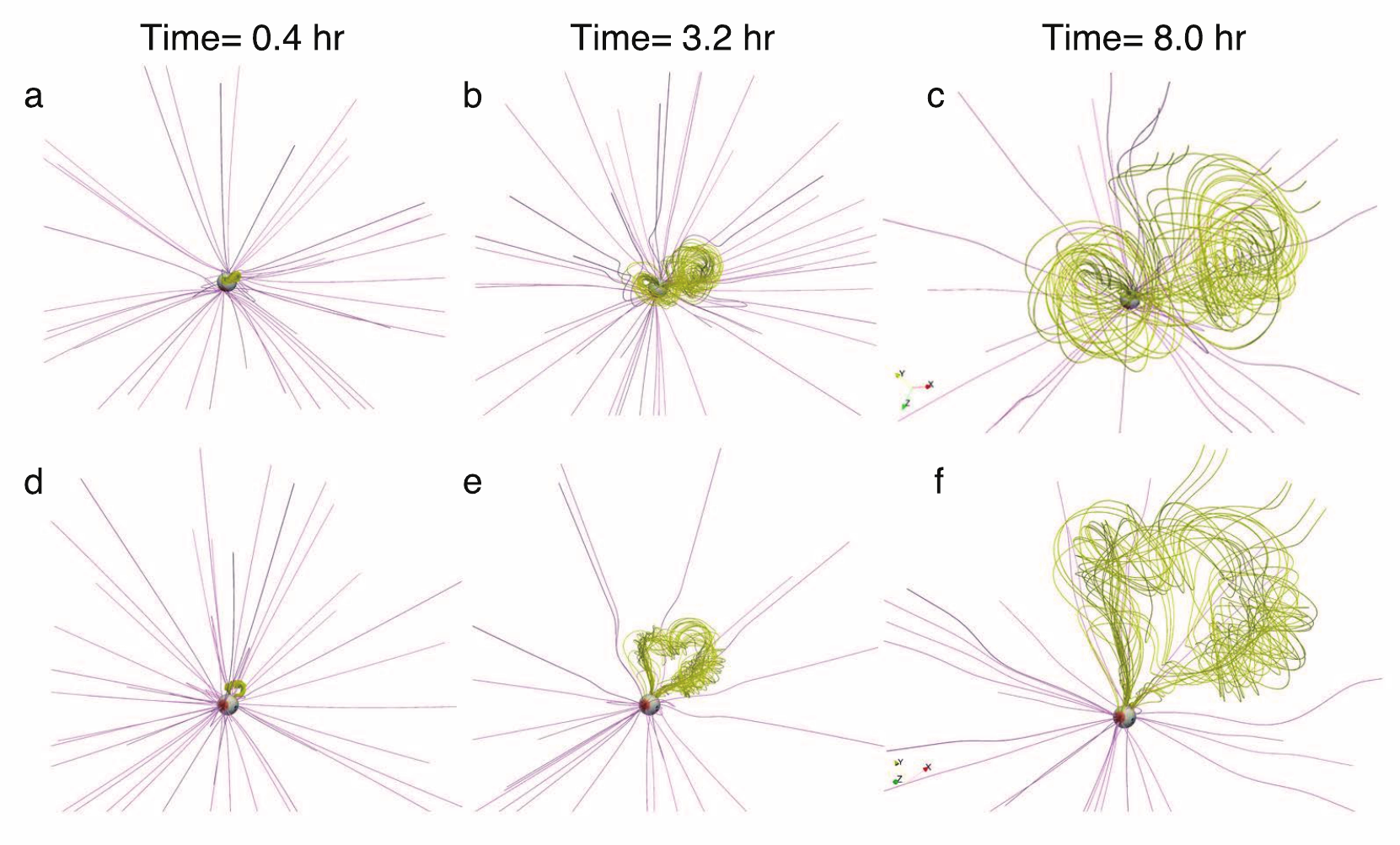}
  \centering
  \caption{Representative magnetic field lines displaying the global evolution of the CME. The yellow and pink tubes represent the magnetic structures of the CME flux rope and the ambient solar wind, respectively. The top row (a, b, c) illustrates the results viewed from the top of the flux rope, while the bottom row (d, e, f) presents the side views. Snapshots at different time intervals are displayed in the left (a, d), middle (b, e), and right (c, f) panels, corresponding to 0.4, 3.2, and 8.0 hours, respectively. \label{fig3}}
\end{figure*}

Figure~\ref{fig3} illustrates the propagation of an S-shaped CME flux rope, where the flux-rope field lines are traced from the footpoints of the initial flux rope. These illustrations effectively capture the significant changes in the overall morphology of the CME flux rope as it propagates outward. It is found that the volume occupied by the CME flux rope expands considerably, which could be attributed to two main factors. Firstly, the magnetic-pressure gradient between the flux rope and the surrounding solar atmosphere can drive the flux rope expansion in all directions \citep{Scolini2019}. Secondly, magnetic reconnection occurring between the legs of overlying field lines leads to the formation of twisted field lines that wrap around the original flux rope, creating a hierarchical structure with varying twist, magnetic connectivity and temperature distributions \citep{Guojh2023}. As shown in Figure~\ref{fig3}f, after 8~hrs, some open field lines with only one line-tied footpoint appear on the periphery of the CME flux rope. Besides, there is the presence of underlying flare loops and highly curved field lines above. These signals strongly indicate the occurrence of magnetic reconnection during the CME propagation. It is worth noting that we use an ideal MHD solver in this paper, which means that magnetic reconnection is due to the numerical diffusion.

\begin{figure*}
  \includegraphics[width=17cm,clip]{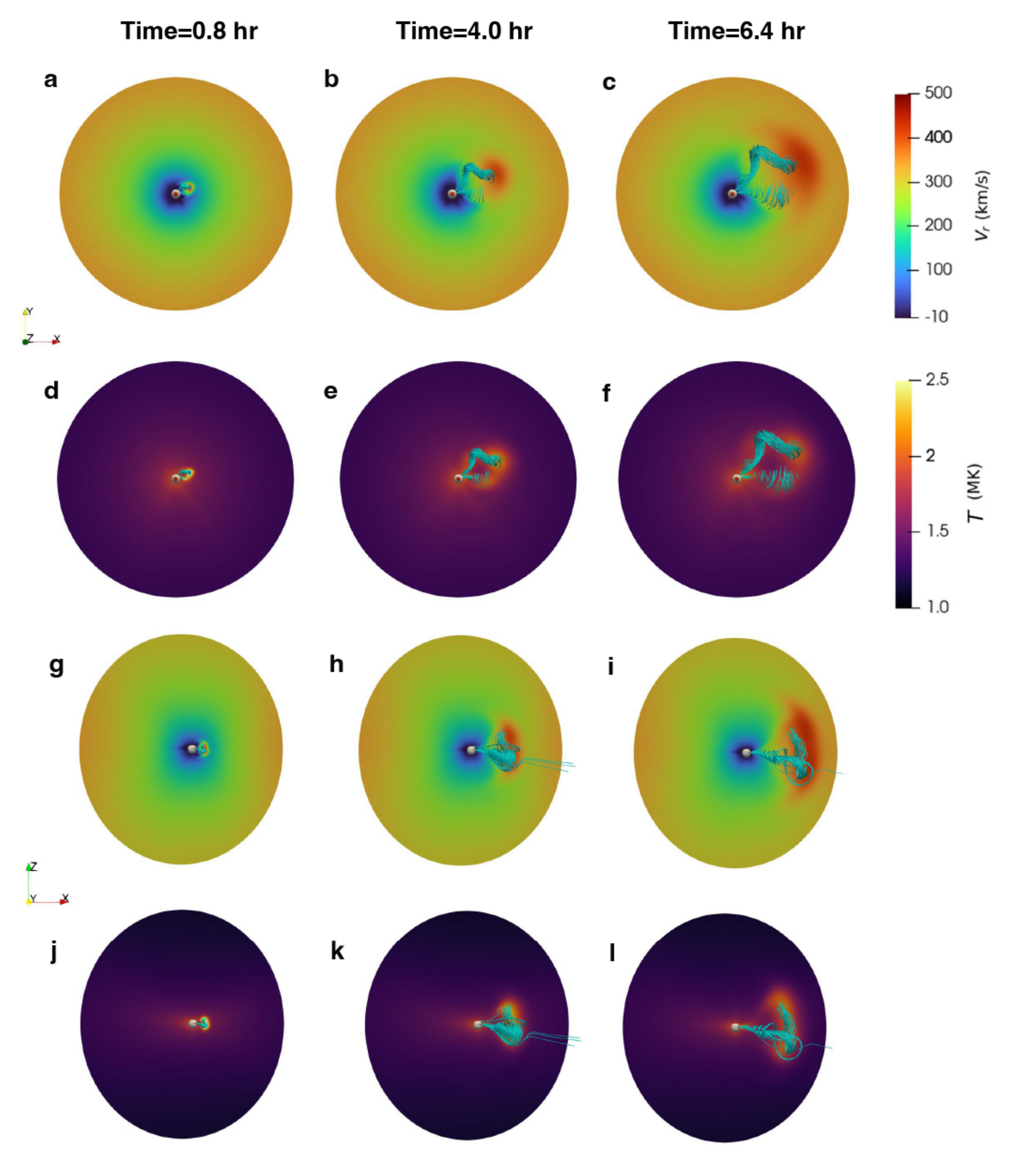}
  \centering
  \caption{Radial velocity and temperature distributions in equatorial and meridian planes. The top (a--f) and bottom two (g--l) panels show the results in equatorial and meridian planes, respectively. Panels~(a)--(c) and (g)--(i) display the distributions of the radial velocity, and panels~(d)--(f) and (j)--(l) correspond to those of the temperature. Cyan tubes are some typical field lines illustrating the CME flux rope. \label{fig4}}
\end{figure*}

Then, we present the velocity and temperature distributions in the equatorial and meridian planes during the CME propagation, as illustrated in Figure~\ref{fig4}. The radial velocity distributions clearly reveal the presence of a bow-shock structure straddling over the CME flux rope, which grows in size as it propagates outward. The intermediate area between the CME flux rope and the leading shock corresponds to the sheath region, which is formed due to plasma compression resulting from the propagation of the flux rope \citep{Kilpua2017, Regnault2020}. Intriguingly, it is noticed that the regions swept by the CME-driven shock do not return to a quasi-steady state prior to the eruption in an elastic manner, and a sustained high-speed flow is induced. This may be attributed to the ongoing outflows resulting from magnetic reconnection processes and the downstream flows after the shock wave. Regarding the temperature slices, it is apparent that a toroidal region experiences significant heating and closely envelops the flux rope, which may be due to the dissipation of the strong magnetic fields within the flux rope itself and the magnetic reconnection that takes place around it. Additionally, the temperature of the flux rope decreases considerably as it propagates outward.

\begin{figure}
  \includegraphics[width=8cm,clip]{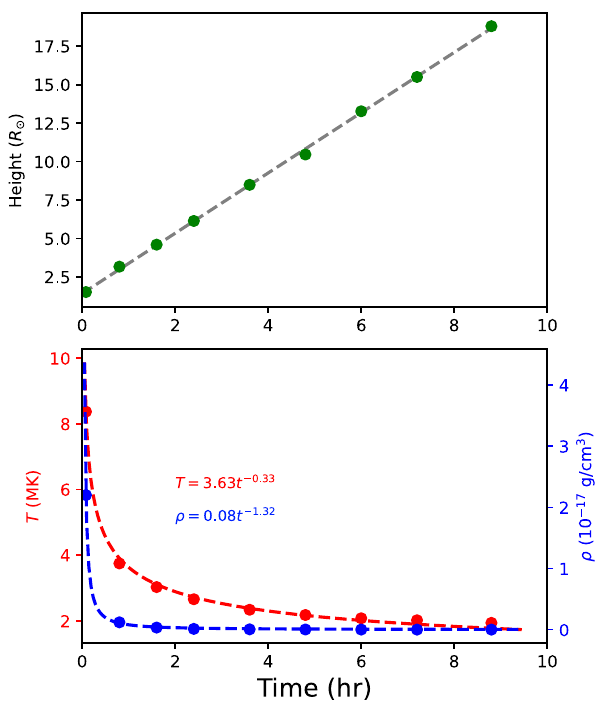}
  \centering
  \caption{Kinetic and thermodynamic evolution of the CME flux rope. Panel~(a) corresponds to the time-height diagram of the CME flux rope. Panel~(b) presents the temperature (red) and density (blue) evolution of the CME flux rope over time.  \label{fig5}}
\end{figure}

Figure~\ref{fig5} exhibits the quantitative results of the kinetics and thermodynamics evolution of the CME flux rope. Panel~(a) displays the time-distance diagram of the CME flux rope, by tracking the highest temperature point along the radial path. Its movement can be linearly fitted with an average speed of 379~km s$^{-1}$, which is in accordance with the speed range of the observed CMEs \citep{Chen2011}. Panel~(b) displays the evolutions of temperature and density, which are fitted with a power-law function of time for each. This reveals that, as the CME flux rope propagates, both its temperature and density undergo decreasing. In addition, the density shows a more rapid decline than the temperature, as inferred from the fitted power exponents. In particular, the CME flux rope is fairly high in temperature within 2$R_{\odot}$, reaching up to 10 MK, which is in line with the hot channels observed in the source regions. Afterwards, both the temperature and density of the CME flux rope decrease rapidly in one hour, and subsequently both become more gradual. By the time the CME flux rope reaches a distance of around 20$R_{\odot}$, its temperature has decreased to 2 MK, and the density has decreased by three orders of magnitude compared to the initial values. Such a decline in the temperature and density of the CME also appeared in the AWSoM simulation \citep{Jin2013} and the PLUTO simulation \citep{Regnault2023}.

\subsection{CME structure and its in-situ measurements}

\begin{figure*}
  \includegraphics[width=16cm,clip]{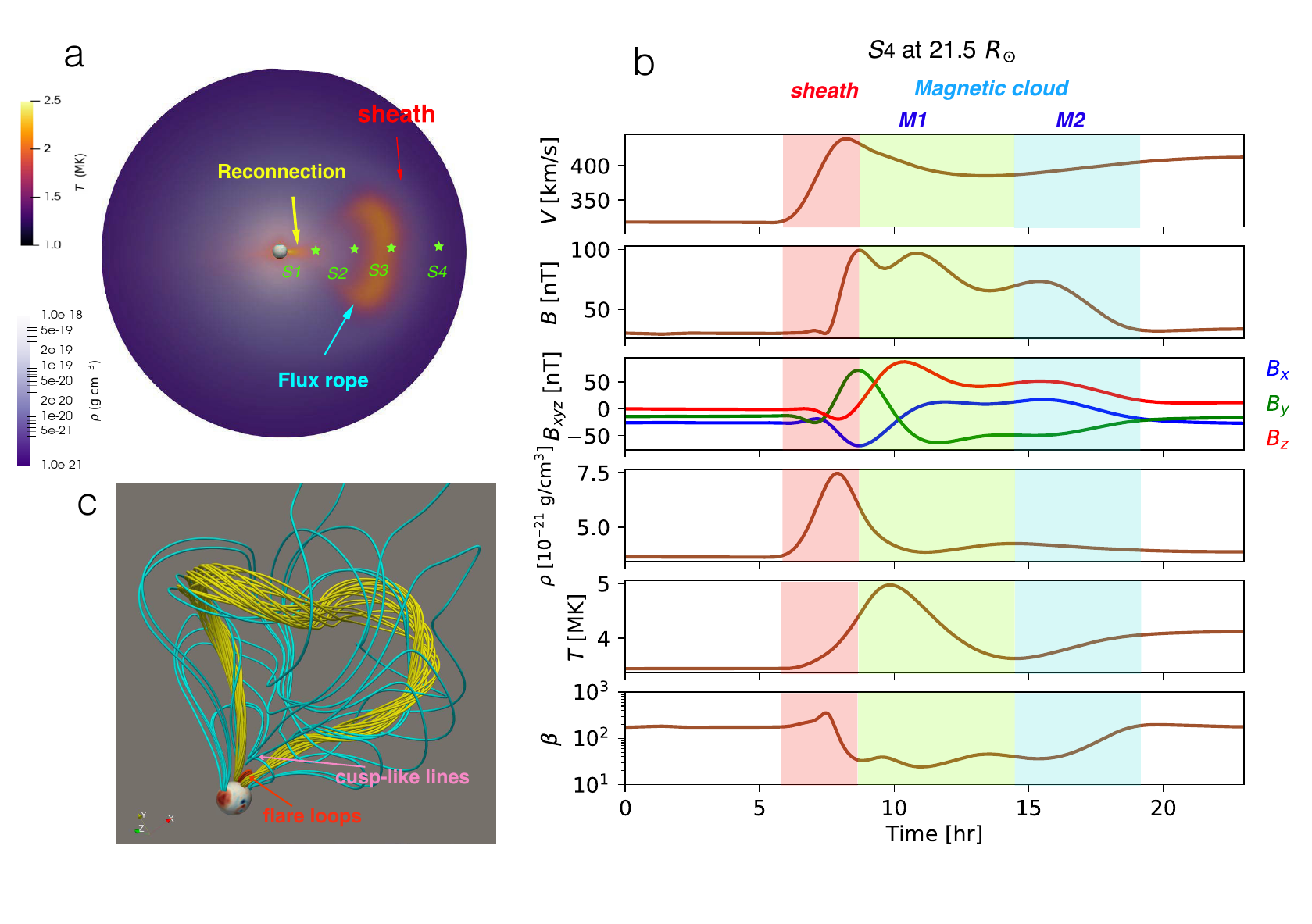}
  \centering
  \caption{Magnetic-field and thermodynamic structure of the CME. Panel~(a) provides a 2D visualization of the density and temperature in the meridian plane. Panel~(b) corresponds to the in-situ plasma profile measured by a virtual satellite at around 21.5$\;R_{\odot}$, where the red, green and blue bands represent the sheath and two sub-parts of the magnetic cloud (M1 and M2), respectively. Panel~(c) illustrates some typical field lines around the CME, where the yellow, cyan and red tubes represent the field lines stem from the original inserting flux rope, newly-formed twisted field lines during the CME propagation and the underlying flare loops, respectively. \label{fig6}}
\end{figure*}

Figure~\ref{fig6}a exhibits the density distribution overlaid by the high-temperature transparent contours ($T>2\;$MK) in the meridian plane. It is found that the corona disturbed by the CME is highly hierarchical, including a leading density enhancement, a following high-temperature magnetic ejection, and a hot tail near the Sun. Such a scenario can be explained by the standard flare model \citep{Carmichael1964, Sturrock1966, Hirayama1974, Kopp1976} as follows. The eruptive flux rope induces the formation of a current sheet below it, in which magnetic reconnection causes positive feedback to the further rising of the flux rope. Meanwhile, the ascending flux rope could produce fast-mode MHD waves propagating upward and result in the plasma compression ahead of it, and thus forms the bright leading front of the CME \citep{Chen2009, Chen2011, Guojh2023}. Therefore, it can be expected that these three distinct regions in Figure~\ref{fig6}a should correspond to the CME leading front, the flux rope, and the reconnection area in the standard flare model, respectively.  
Hereafter, following the approach to analyze the interplanetary CMEs (ICMEs), we plot the in-situ plasma profiles measured by a virtual spacecraft at point S4 with a radial distance of 21.5$R_{\odot}$ in Figure~\ref{fig6}c (this radial distance serves as the inner boundary of the EUHFORIA simulations). One can recognize the arrival of the CME from the plasma profiles, and its resulting disturbances on the solar wind can be divided into the following parts. In the leading part (red band in Figure~\ref{fig6}b), the temperature, velocity, density, and plasma $\beta$ increase a lot. Besides, there is also a prominent fluctuation in the magnetic fields. The above features are generally identified as the sheath formed ahead of the magnetic cloud \citep{Kilpua2017, Regnault2020, Linan2023}. It is worth noting that the variations of the magnetic-field components in our simulation are different from those modeled by the TDm flux rope \citep{Linan2023} despite the background solar wind is the same. This demonstrates that the initial flux rope could also influence the sheath structure of its resulting CME. 

The area in the wake of the sheath exhibits several characteristic features commonly observed in interplanetary magnetic clouds, i.e., an enhanced magnetic-field strength, smoothly changing magnetic-field orientation, low plasma $\beta$, and a decreasing in velocity. These signatures are consistent with the main body of the CME, namely, the helical flux rope. Moreover, this can be further divided into two sub-parts, referred to as M1 (green band in Figure~\ref{fig6}b) and M2 (blue band in Figure~\ref{fig6}b). In part M1, the vector magnetic fields exhibit the rotation features:~the $B_{x}$ component rotates from the negative to positive, the $B_{y}$ component rotates from the positive to negative, and the $B_{z}$ component almost remains positive. This orientation reflects the right handed helix with positive magnetic helicity, which is consistent with that of the initial flux rope. In contrast to part M1, the magnetic-field orientation of part M2 almost remains unchanged throughout. Moreover, the solar-wind velocity and temperature in part M2 start to increase. We speculate that such a secondary acceleration and heating is the consequence of magnetic reconnection, corresponding to hot reconnection outflows. Similar in-situ plasma profiles of CMEs detected near the Sun were also found in previous COCONUT-CME simulations \citep{Linan2023}, the detection of the Parker Solar Probe at around 57.4$R_{\odot}$ \citep{Korreck2020} and the observations derived from the radio occultation measurements with the Akatsuki spacecraft \citep{Ando2015}. 

It should be pointed out that the criteria employed to distinguish the boundaries between these distinct regions are described as follows:~(1) the sheath is principally identified by a prominent bump in the plasma $\beta$ profile, accompanied by the elevated velocity and density;~(2) the magnetic cloud is primarily characterized by the plasma $\beta$ profile, which exhibits a significantly lower value compared to that of the background solar wind;~(3) the transition point from M1 to M2 is selected as the presence of secondary acceleration and heating, which can serve as an indicator of the newly formed flux rope resulting from magnetic reconnection. Specifically, this secondary increase in velocity and temperature profiles, along with the presence of newly formed twisted field lines due to reconnection, have also been seen in CMEs initiated from TDm flux ropes \citep{Linan2023}. To further illustrate the magnetic structure of the ejection, we plot some typical field lines around the CME in Figure~\ref{fig6}c. One can see two types of twisted field lines with different connectivity: in the first category represented by yellow tubes, the flux rope shows a coherent structure and has an analogous configuration with the inserted flux rope, implying that it is likely to stem from the eruption of the preexisting flux rope. However, for the second type (cyan tubes), the magnetic connectivity is significantly different from that of the former, manifested as the different footpoints and twist characteristics, which suggests that they could be formed as a result of non-ideal processes, such as magnetic reconnection. In addition, we can also recognize the cusp-like highly-curved field lines and the flare loops below (as shown in Figure~\ref{fig6}c), which is the strong evidence for the occurrence of the magnetic reconnection during the CME propagation.

\begin{figure*}
  \includegraphics[width=\textwidth,clip]{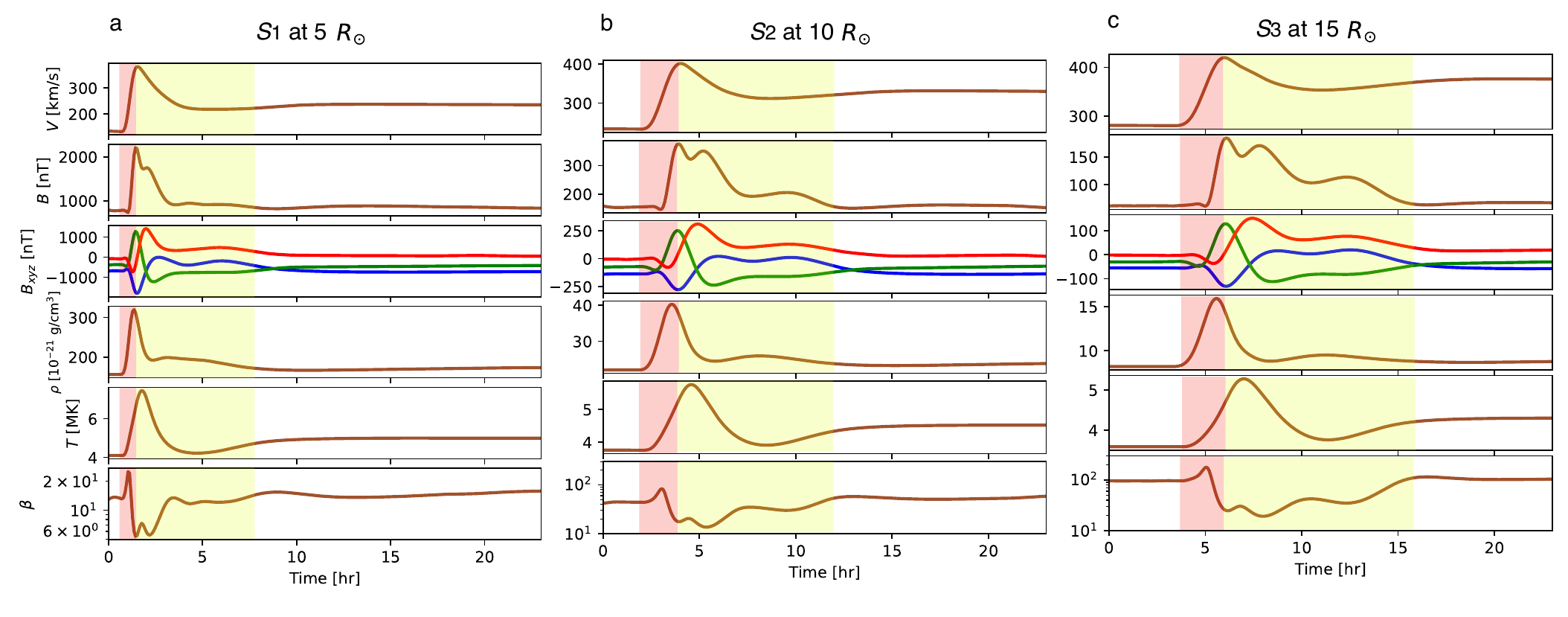}
  \centering
  \caption{In-situ plasma profiles that are measured at different distances from the Sun. Panels~(a), (b), and (c) present the results measured by S1 (5$\;R_{\odot}$), S2 (10$\;R_{\odot}$) and S3 (15$\;R_{\odot}$) marked in Figure~\ref{fig6}a, respectively. The red and yellow bands represent the sheath and magnetic cloud, respectively. \label{fig7}}
\end{figure*}

To investigate the temporal evolution of the CME during its propagation, we analyze the localized plasma profiles at different distances from the Sun, which are depicted in Figure~\ref{fig7}, specifically at S1 (5$R_{\odot}$), S2 (10$R_{\odot}$), and S3 (15$R_{\odot}$). Firstly, the volume of the CME body enlarges a lot as it propagates outward (yellow band), which is attributed to its self-expansion and injection of the newly twisted fluxes due to magnetic reconnection. One effect of the CME expansion is the decline of its magnetic-field strength ($B$). At 15$R_{\odot}$, the strength decreases by a factor of ten compared to that at 5$R_{\odot}$. Moreover, magnetic reconnection during the propagation can cause restructuring for the original flux rope, resulting in a more intricate magnetic structure. In the 3D illustration of the magnetic field lines, this can be manifested as the twisted field lines with different connectivity and footpoints, as shown in Figure~\ref{fig6}c. In one-dimensional in-situ plasma profiles, this can be reflected in the formation of a series of sub-peaks inside the magnetic cloud. Around 5$R_{\odot}$, only one single prominent peak is observed in the $B$ profile. However, as the CME reaches approximately 15$R_{\odot}$, its profile evolves into a structure consisting of multiple peaks. Regarding the velocity profile, it exhibits continuous increase when propagating, although the slope of the leading front becomes flatter compared to earlier stages \citep{Jin2013}.

\section{Discussion} \label{sec:dis}

\subsection{Complexity of the magnetic topology of CME flux ropes}
It is widely accepted that magnetic flux ropes are commonly present within CMEs and play a crucial role in explaining CME generation and the observational characteristics of the subsequent interplanetary counterparts \citep{Chen2011}. As a result, the prevailing approach in most existing numerical prediction models is to initiate a CME by inserting an eruptive magnetic flux rope. One of the simple methods to achieve this is by incorporating an analytical (or semi-analytical) flux rope with a coherent structure and simple morphology at the inner boundary of the heliosphere, such as around 21.5$R_{\odot}$ \citep{Verbeke2019, Scolini2019, Scolini2020, Maharana2021, Maharana2023}. However, several issues were ignored in these simulations:~(1) can the CME flux rope situated at approximately 20$R_{\odot}$ still be regarded as a coherent structure? (2) is it possible that the physical processes occurring in the corona can lead to a deviation of the flux-rope topological structure from that prior to the eruption?

The majority of observations and numerical simulations indicated that the flux rope would undergo drastic magnetic reconnection during eruption, which will significantly change its magnetic structure \citep{Liur2020}. For example, \citet{Wangws2017} demonstrated that a flux rope can be formed due to the reconnection in the eruption process rather than existing prior to the eruption. \citet{Gou2023} claimed that the magnetic reconnection in the eruption process can lead to a complete replacement for the flux of the original flux rope. Furthermore, the detection of superthermal electrons detected in the interplanetary space indicated that the magnetic cloud may be composed of the closed and open field lines \citep{Gosling1995, Crooker2004}. To explain the complex phenomena in observations, \citet{Aulanier2019} proposed a new 3D flare model consisting of three types of magnetic reconnection geometries, i.e., the reconnection in the overlying field lines (called aa--rf type), the reconnection between the flux rope and the ambient arcades (ar--rf type), and the reconnection in the flux rope field lines (rr--rf type). Recently, \citet{Guojh2023} performed a data-driven MHD simulation, and demonstrated that the magnetic reconnection in the eruption can lead to a change in the temperature, orientation, and twist number of the flux rope. Therefore, it is expected that magnetic reconnection would result in the CME flux rope deviating from the coherent structure when it reaches the interplanetary space.

\begin{figure*}
  \includegraphics[width=18cm,clip]{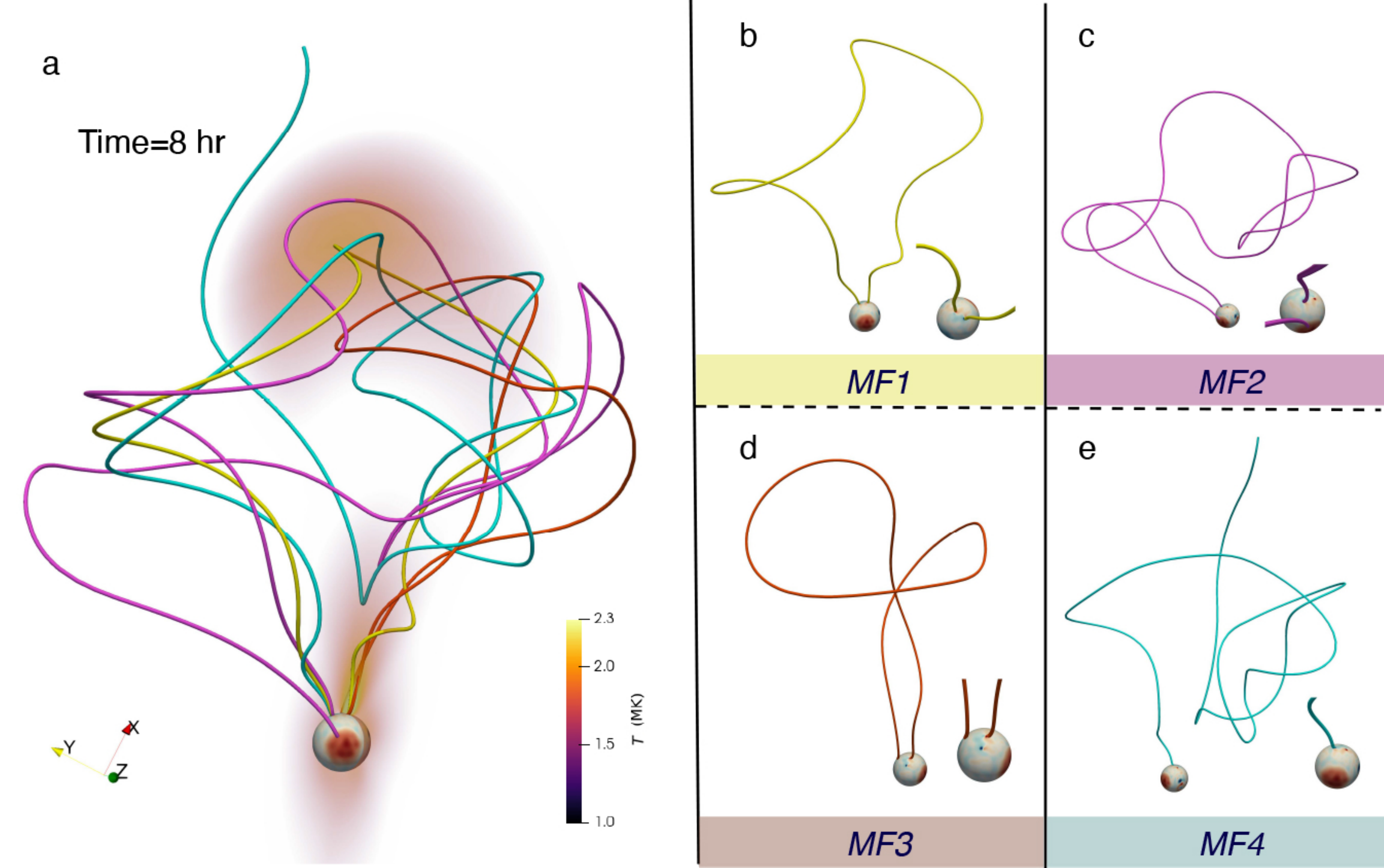}
  \centering
  \caption{Magnetic structure of the CME flux rope at around 20$\;R_{\odot}$. Panel~(a) showcases a collection of typical field lines of the CME flux rope, which are enveloped by a high-temperature transparent contour. Panels (b)--(d) illustrate the field lines in panel~(a) individually, providing details into their magnetic connectivity and the footpoint placements. \label{fig8}}
\end{figure*}

To reveal the magnetic topology of the CME flux rope at around 20$R_{\odot}$, we examine some representative twisted field lines traced from the flux-rope footpoints, which are shown in Figure~\ref{fig8}. It is evident that the magnetic structure of the CME at 20$R_{\odot}$ differs from its progenitor in the source region. When the CME flux rope reaches around 20$R_{\odot}$, it is composed of a mix of closed and open field lines. The right panels in Figure~\ref{fig8} illustrate the connectivity of these twisted field lines and characterize them individually. They exhibit notable distinctions in terms of the footpoint placements and orientation. For example, the footpoints of MF1 remain attached to the same locations as the initial flux rope. However, for MF2 and MF3, one of their footpoints migrates toward the polar regions, suggesting the occurrence of large-scale interchange reconnection between the flux-rope field lines and the neighbouring open field lines extending from the polar regions. Additionally, we also identify the open field line (MF4) with only one footpoint tied to the Sun, which may be formed due to the reconnection between the flux-rope field lines and adjacent open streamers. 

It is worth noting that, we can see a secondary peak of the in-situ velocity and density profiles inside the magnetic cloud characterized by the low-$\beta$ region compared to the background solar wind. Intriguingly, \citet{Ando2015} observed a similar secondary enhancement of the velocity and density profiles at a distance of 12.7~$R_{\odot}$, which was derived from the radio occultation measurements with the Akatsuki spacecraft \citep{Nakamura2011}. Drawing from the insights of the numerical model performed by \citet{Shiota2005}, they suggested that fast flows due to magnetic reconnection are responsible for this secondary velocity and density peak inside the CME. Our simulation results are in high agreement with their observations. Both the simulation and observation indicate that the second enhancement in velocity and density inside the CME could serve as a signature of plasma outflows of magnetic reconnection. Moreover, magnetic reconnection could also contribute to the complexity of the magnetic structure of the CME, as depicted in Figures~\ref{fig6}c and \ref{fig8}.

Accordingly, the magnetic topology of the CME flux rope is considerably complicated and deviates from a coherent structure. First of all, the footpoints of a CME flux rope may be more than two and are not always fixed to the initial locations. Additionally, the CME flux rope is not always composed of closed field lines anchored to the solar surface. The interchange reconnection between the CME flux rope and open streamers can form open twisted field lines \citep{Masson2013}. A similar topology has also been found in a data-constrained simulation performed by \citet{Lugaz2011}. They found that the reconnection with the ambient coronal holes can yield open field lines inside the CME. These open field lines are usually used to explain the impulsive SEP bursts \citep{Cane1986, Masson2013} and type \uppercase\expandafter{\romannumeral3} radio burst \citep{Krucker2011, Chen2018}, in which the energetic particles accelerated in the flare are released into the interplanetary space along the open field lines. Therefore, our simulation results indicate that the approach to input a CME at 20$R_{\odot}$ with an analytical model is risky for the space weather prediction, which is likely to lead to a failure in reproducing some ICME events exhibiting complex magnetic-field profiles and associated with impulsive SEP events. Hence, there is a need to develop models that couple the solar corona and interplanetary space, as done in some works \citep{Jin2012, Jin2017, Zhou2017, Torok2018}. This approach holds the potential to provide more self-consistent CME models, leading to a notable improvement in the prediction accuracy.

\subsection{Importance of solar poles on global coronal modelings}
Another noteworthy point is the importance of the solar poles. As illustrated in Figure~\ref{fig8}, the footpoints of some CME flux-rope field lines (MF2 and MF3) drift to the poles due to magnetic reconnection. This demonstrates that the streamers extending from the solar poles stand a good chance to have interaction with a CME even though it is initiated from the equator. However, owing to the singularity problem of the solar poles in the spherical coordinates, the majority of global MHD simulations still fail to contain the solar poles. \citet{Perri2023} investigated the impacts of the input magnetograms and solar poles on the solar wind fields, and found that the filling of poles is fairly important even for the forecasts concentrated on the ecliptic plane. 

Our simulation indicated that the filling of solar poles not only influences the background solar wind, but also impacts the structures of CMEs occurring in this medium: the CME flux rope may reconnect with the open field lines extending from the poles. This suggests that the inclusion of solar poles is indispensable for the global space-weather prediction-aimed models.

\section{Summary} \label{sec:sum}

In this paper, we presented the implementation of the RBSL magnetic flux rope in COCONUT, and then modeled the self-consistent propagation of an S-shaped flux rope from the solar surface to 25$R_{\odot}$ with the numerical model. We made an analysis of the kinematics, magnetic topology, and in-situ plasma properties of the modeled CME. The results demonstrate the potential of the RBSL flux rope in reproducing the CMEs that are more in line with observations: this is bound to play an important role in future space weather forecasting. To summarize, our simulation leads to the following main results:

\begin{enumerate}
\item{We implemented the RBSL flux rope model in COCONUT, which is an advanced method to construct the CME pioneer that is strongly constrained by observations. This allows to construct a flux rope with an axis of an arbitrary path. As such, we can model a flux rope where its axis path is directly measured from the proxies of the observed flux ropes, such as the filaments, sigmoids, and hot channels. We believe that the RBSL technique has great potential in future event studies and more accurate space-weather predictions.}

\item{The magnetic structure of the flux rope of a CME at around 20$R_{\odot}$ may deviate from the coherent structure. We found that magnetic reconnection can significantly change the magnetic topology of the CME flux rope. For example, the reconnection between the legs of the overlying field lines can inject newly-formed twisted field lines encircling the preexisting flux rope, which may lead to restructuring of the original flux rope. Besides, the interchange reconnection occurring between the flux rope and the open streamers can result in some open twisted field lines. Additionally, the reconnection between the flux rope and ambient sheared arcades can cause the flux-rope footpoint to drift. Therefore, the CME flux rope at around 20$R_{\odot}$ is a mix of open and closed field lines with far apart footpoints.}

\item{Solar poles are indispensable in the global coronal models aimed at the CME predictions. We found that there exists magnetic reconnection between the CME flux rope and open field lines extending from the polar regions, even though the initial flux rope is launched at the equatorial plane. Due to the advantage of the unstructured mesh grids in COCONUT, the polar singularity can be averted naturally.}
\end{enumerate}

Based on the advantage of the RBSL flux rope, the next step of our work could be to model a truly observed CME event. Compared to our previously implemented TDm flux rope model in COCONUT \citep{Linan2023}, where its axis is fixed as a toroidal shape, the RBSL flux rope exhibits greater flexibility (arbitrary path), leading to a better resemblance to the observations. As demonstrated in our previous observational data-driven MHD simulations constrained in the local Cartesian coordinates \citep{Guoy2021, Guojh2023}, the onset process of the CME resulting from the RBSL flux rope coincides well with the observations. Following these, we will perform the data-driven simulation with COCONUT to model a truly observed CME in a global domain, which can be used to study the propagation of the global extreme-ultraviolet waves \citep{Chen2016}, nature of the CME three-part structure \citep{Chen2011, Song2022}, and large-scale current sheet \citep{Lin2005, Cheng2018}. Moreover, to make the simulation results more comparable with observations, we also plan to conduct the forward modeling with the 3D data of the COCONUT simulation in future works, such as the synthesis of the white-light and extreme ultraviolet (EUV) radiations. In addition, the CME at around 21.5$R_{\odot}$ obtained from the COCONUT simulations can also be adopted as the input of the EUHFORIA simulations. It is expected that the CME undergoing the self-consistent evolution in the corona would yield higher prediction accuracy compared to the method that launches an analytical CME model with a simple shape at the inner boundary of the interplanetary space. 

It is evident that the current version still has several drawbacks that need to be tackled. For example, the plasma $\beta$ distribution in our simulation is larger than that in observations \citep{Korreck2020}. The higher plasma $\beta$ compared to the observations could be attributed to the quasi-isothermal approximation in the polytropic model, as mentioned in \citet{Regnault2023}. This approximation may lead to the following impacts on our results. First, expanding CMEs require pushing against the higher pressure from the ambient solar wind plasma in a high plasma $\beta$ environment, potentially resulting in less expansion compared to observations. Second, the $\rm Alfv\acute{e}n$ speed and Lorentz force become lower compared to the sound speed and pressure gradient force in a high plasma $\beta$ environment, thereby diminishing the effects of magnetic reconnection on CME acceleration. Moreover, the parametric survey conducted by \citet{Ni2012} indicated that plasma $\beta$ also affects the critical Lundquist number for the plasmoid instability to occur. Consequently, the intricate magnetic topology exhibited in our simulation may have possible relevance to the modeling setup. Nevertheless, some works have demonstrated that the influences of plasma $\beta$ may still be limited. For instance, \citet{Ni2012} found that global evolution and the reconnection rate are very similar in the temperature-stratified atmosphere even when plasma $\beta$ changes from 0.2 to 50. Additionally, \citet{Wang2021} found that the probability distribution functions of magnetic islands slightly vary for the cases with different plasma $\beta$ values. \citet{Linton2002} studied the reconnection process of two twisted flux ropes and found that magnetic energy of the post-reconnection equilibrium state exhibits qualitative similarity in low-$\beta$ and high-$\beta$ cases. The CME models in a more realistic atmosphere will be developed in our future endeavors. In particular, \citet{Brchnelova2023} recently developed a new method to achieve the more realistic plasma $\beta$ value in the reconstruction of solar wind.

It is noted in passing that the center of the CME flux rope undergoes excessive heating during the initiation process, reaching up to 10~MK within the first five minutes, followed by a gradual decrease, as illustrated in Figure~\ref{fig5}. This phenomenon may be due to the following two factors. On the one hand, the eruption in our model does not start from an equilibrium flux rope. Albeit the force-free condition is automatically fulfilled inside the RBSL flux rope, the external equilibrium condition between the flux rope and the background magnetic fields is not achieved, where the intensity of the electric current flowing inside the flux rope exceeds the equilibrium value derived from Shafranov’s equation. As such, the flux rope will immediately rise when the simulation is advanced forward in time, without establishing a thermodynamic coupling with the ambient solar wind. Consequently, additional heating may be induced compared to eruptions originating from stable flux ropes in the realistic corona. On the other hand, the grid resolution is too coarse to accurately resolve the thermodynamic evolution of the CME flux rope in the low corona with strong magnetic fields, thereby contributing to numerical heating. It should be pointed out that \citet{Jin2013} found that the two-temperature model (coupling the thermodynamics of electron and proton populations) is significantly better in accurately reproducing the CME temperature. To sum up, the thermodynamic evolution of the CME flux rope in the current version remains considerably simplified and departs from that in the realistic corona. For instance, many nonadiabatic terms in the energy equation are omitted in the energy equation of the polytropic model, such as thermal conduction and radiation losses. Moreover, to counterbalance the radiation losses and drive the fast solar wind, the additional physical heating term must also be taken into account. The dissipation of Alfv\'en waves was demonstrated to be a potential candidate and has been implemented in some global coronal models \citep{VanderHolst2010, Mikic2018}, which can reproduce the EUV and white-light emissions very well, and drive the fast solar wind. Such a radiation MHD model should provide a better overall description for the CME propagation, particularly in terms of the thermodynamic evolution, than the currently employed polytropic model. 

Apart from the thermodynamic process, the initial magnetic-field model is also an important factor to influence the relaxed solar wind. So far, the simulations performed by COCONUT used the PFSS model as the initial magnetic field. However, there is a well-accepted fact that the coronal magnetic fields deviate from a potential state, which is generally approximated by the nonlinear force-free fields \citep[NLFFF;][]{Wiegelmann2021}. Hence, it is speculated that the adoption of the global NLFFF model \citep{Guo2016b, Yeats2016, Koumtzis2023} could be more beneficial in speeding-up the convergence. All in all, we believe that these future updates should hold significant promise in facilitating more precise and timely space weather forecasting.

\begin{acknowledgements}
 The SDO data are available courtesy of NASA/SDO and the AIA and HMI science teams. This work is supported by the National Key R\&D Program of China (2020YFC2201200, 2022YFF0503004), NSFC (12127901), projects C14/19/089  (C1 project Internal Funds KU Leuven), G.0B58.23N  (FWO-Vlaanderen), SIDC Data Exploitation (ESA Prodex-12), and Belspo project B2/191/P1/SWiM. J.H.G. was supported by the China Scholarship Council under file No.\ 202206190140. The resources and services used in this work were provided by the VSC (Flemish Supercomputer Centre), funded by the Research Foundation - Flanders (FWO) and the Flemish Government. 
\end{acknowledgements}

\bibliographystyle{aa}
\bibliography{ms}

\begin{thebibliography}{116}
\expandafter\ifx\csname natexlab\endcsname\relax\def\natexlab#1{#1}\fi

\bibitem[{{Ando} {et~al.}(2015){Ando}, {Shiota}, {Imamura}, {Tokumaru}, {Asai},
  {Isobe}, {P{\"a}zold}, {H{\"a}usler}, \& {Nakamura}}]{Ando2015}
{Ando}, H., {Shiota}, D., {Imamura}, T., {et~al.} 2015, Journal of Geophysical
  Research (Space Physics), 120, 5318

\bibitem[{{Asvestari} {et~al.}(2022){Asvestari}, {Rindlisbacher}, {Pomoell}, \&
  {Kilpua}}]{Asvestari2022}
{Asvestari}, E., {Rindlisbacher}, T., {Pomoell}, J., \& {Kilpua}, E. K.~J.
  2022, \apj, 926, 87

\bibitem[{{Aulanier} \& {Dud{\'\i}k}(2019)}]{Aulanier2019}
{Aulanier}, G. \& {Dud{\'\i}k}, J. 2019, \aap, 621, A72

\bibitem[{{Aulanier} {et~al.}(2010){Aulanier}, {T{\"o}r{\"o}k}, {D{\'e}moulin},
  \& {DeLuca}}]{Aulanier2010}
{Aulanier}, G., {T{\"o}r{\"o}k}, T., {D{\'e}moulin}, P., \& {DeLuca}, E.~E.
  2010, \apj, 708, 314

\bibitem[{{Baratashvili} {et~al.}(2022){Baratashvili}, {Verbeke}, {Wijsen}, \&
  {Poedts}}]{Baratashvili2022}
{Baratashvili}, T., {Verbeke}, C., {Wijsen}, N., \& {Poedts}, S. 2022, \aap,
  667, A133

\bibitem[{{Boe} {et~al.}(2020){Boe}, {Habbal}, \& {Druckm{\"u}ller}}]{Boe2020}
{Boe}, B., {Habbal}, S., \& {Druckm{\"u}ller}, M. 2020, \apj, 895, 123

\bibitem[{{Brchnelova} {et~al.}(2022){Brchnelova}, {Ku{\'z}ma}, {Perri},
  {Lani}, \& {Poedts}}]{Brchnelova2022b}
{Brchnelova}, M., {Ku{\'z}ma}, B., {Perri}, B., {Lani}, A., \& {Poedts}, S.
  2022, \apjs, 263, 18

\bibitem[{{Brchnelova} {et~al.}(2023){Brchnelova}, {Ku{\'z}ma}, {Zhang},
  {Lani}, \& {Poedts}}]{Brchnelova2023}
{Brchnelova}, M., {Ku{\'z}ma}, B., {Zhang}, F., {Lani}, A., \& {Poedts}, S.
  2023, arXiv e-prints, arXiv:2306.08874

\bibitem[{Brchnelova {et~al.}(2022)Brchnelova, Zhang, Leitner, Perri, Lani, \&
  Poedts}]{Brchnelova2022}
Brchnelova, M., Zhang, F., Leitner, P., {et~al.} 2022, Journal of Plasma
  Physics, 88, 905880205

\bibitem[{{Burlaga} {et~al.}(1981){Burlaga}, {Sittler}, {Mariani}, \&
  {Schwenn}}]{Burlaga1981}
{Burlaga}, L., {Sittler}, E., {Mariani}, F., \& {Schwenn}, R. 1981, \jgr, 86,
  6673

\bibitem[{{Cane} {et~al.}(1986){Cane}, {McGuire}, \& {von
  Rosenvinge}}]{Cane1986}
{Cane}, H.~V., {McGuire}, R.~E., \& {von Rosenvinge}, T.~T. 1986, \apj, 301,
  448

\bibitem[{{Carmichael}(1964)}]{Carmichael1964}
{Carmichael}, H. 1964, in NASA Special Publication, Vol.~50, 451

\bibitem[{{Chen} {et~al.}(2018){Chen}, {Yu}, {Battaglia}, {Farid}, {Savcheva},
  {Reeves}, {Krucker}, {Bastian}, {Guo}, \& {Tassev}}]{Chen2018}
{Chen}, B., {Yu}, S., {Battaglia}, M., {et~al.} 2018, \apj, 866, 62

\bibitem[{{Chen}(2009)}]{Chen2009}
{Chen}, P.~F. 2009, \apjl, 698, L112

\bibitem[{{Chen}(2011)}]{Chen2011}
{Chen}, P.~F. 2011, Living Reviews in Solar Physics, 8, 1

\bibitem[{{Chen}(2016)}]{Chen2016}
{Chen}, P.~F. 2016, Washington DC American Geophysical Union Geophysical
  Monograph Series, 216, 381

\bibitem[{{Cheng} {et~al.}(2017){Cheng}, {Guo}, \& {Ding}}]{Cheng2017}
{Cheng}, X., {Guo}, Y., \& {Ding}, M. 2017, Science China Earth Sciences, 60,
  1383

\bibitem[{{Cheng} {et~al.}(2018){Cheng}, {Li}, {Wan}, {Ding}, {Chen}, {Zhang},
  \& {Liu}}]{Cheng2018}
{Cheng}, X., {Li}, Y., {Wan}, L.~F., {et~al.} 2018, \apj, 866, 64

\bibitem[{{Cheng} {et~al.}(2013){Cheng}, {Zhang}, {Ding}, {Liu}, \&
  {Poomvises}}]{Cheng2013}
{Cheng}, X., {Zhang}, J., {Ding}, M.~D., {Liu}, Y., \& {Poomvises}, W. 2013,
  \apj, 763, 43

\bibitem[{{Crooker} {et~al.}(2004){Crooker}, {Forsyth}, {Rees}, {Gosling}, \&
  {Kahler}}]{Crooker2004}
{Crooker}, N.~U., {Forsyth}, R., {Rees}, A., {Gosling}, J.~T., \& {Kahler},
  S.~W. 2004, Journal of Geophysical Research (Space Physics), 109, A06110

\bibitem[{{Dedner} {et~al.}(2002){Dedner}, {Kemm}, {Kr{\"o}ner}, {Munz},
  {Schnitzer}, \& {Wesenberg}}]{Dedner2002}
{Dedner}, A., {Kemm}, F., {Kr{\"o}ner}, D., {et~al.} 2002, Journal of
  Computational Physics, 175, 645

\bibitem[{{Feng}(2020)}]{Feng2020}
{Feng}, X. 2020, {Magnetohydrodynamic Modeling of the Solar Corona and
  Heliosphere}

\bibitem[{{Feng} {et~al.}(2007){Feng}, {Zhou}, \& {Wu}}]{Feng2007}
{Feng}, X., {Zhou}, Y., \& {Wu}, S.~T. 2007, \apj, 655, 1110

\bibitem[{{Gibson} \& {Low}(1998)}]{Gibson1998}
{Gibson}, S.~E. \& {Low}, B.~C. 1998, \apj, 493, 460

\bibitem[{{Golub} {et~al.}(2007){Golub}, {DeLuca}, {Austin}, {Bookbinder},
  {Caldwell}, {Cheimets}, {Cirtain}, {Cosmo}, {Reid}, {Sette}, {Weber},
  {Sakao}, {Kano}, {Shibasaki}, {Hara}, {Tsuneta}, {Kumagai}, {Tamura},
  {Shimojo}, {McCracken}, {Carpenter}, {Haight}, {Siler}, {Wright}, {Tucker},
  {Rutledge}, {Barbera}, {Peres}, \& {Varisco}}]{Golub2007}
{Golub}, L., {DeLuca}, E., {Austin}, G., {et~al.} 2007, \solphys, 243, 63

\bibitem[{{Gosling}(1993)}]{Gosling1993}
{Gosling}, J.~T. 1993, \jgr, 98, 18937

\bibitem[{{Gosling} {et~al.}(1995){Gosling}, {Birn}, \& {Hesse}}]{Gosling1995}
{Gosling}, J.~T., {Birn}, J., \& {Hesse}, M. 1995, \grl, 22, 869

\bibitem[{Gou {et~al.}(2023)Gou, Liu, Veronig, Zhuang, Li, Wang, Xu, \&
  Wang}]{Gou2023}
Gou, T., Liu, R., Veronig, A.~M., {et~al.} 2023, Nature Astronomy

\bibitem[{{Gui} {et~al.}(2011){Gui}, {Shen}, {Wang}, {Ye}, {Liu}, {Wang}, \&
  {Zhao}}]{Gui2011}
{Gui}, B., {Shen}, C., {Wang}, Y., {et~al.} 2011, \solphys, 271, 111

\bibitem[{{Guo} {et~al.}(2023{\natexlab{a}}){Guo}, {Qiu}, {Ni}, {Guo}, {Li},
  {Gao}, {Schmieder}, {Poedts}, \& {Chen}}]{Guojh2023b}
{Guo}, J., {Qiu}, Y., {Ni}, Y., {et~al.} 2023{\natexlab{a}}, arXiv e-prints,
  arXiv:2308.08831

\bibitem[{{Guo} {et~al.}(2021{\natexlab{a}}){Guo}, {Ni}, {Qiu}, {Zhong}, {Guo},
  \& {Chen}}]{Guojh2021}
{Guo}, J.~H., {Ni}, Y.~W., {Qiu}, Y., {et~al.} 2021{\natexlab{a}}, \apj, 917,
  81

\bibitem[{{Guo} {et~al.}(2023{\natexlab{b}}){Guo}, {Ni}, {Zhong}, {Guo}, {Xia},
  {Li}, {Poedts}, {Schmieder}, \& {Chen}}]{Guojh2023}
{Guo}, J.~H., {Ni}, Y.~W., {Zhong}, Z., {et~al.} 2023{\natexlab{b}}, \apjs,
  266, 3

\bibitem[{{Guo} {et~al.}(2023{\natexlab{c}}){Guo}, {Guo}, {Ni}, {Ding}, {Chen},
  {Xia}, {Keppens}, \& {Yang}}]{Guoy2023}
{Guo}, Y., {Guo}, J., {Ni}, Y., {et~al.} 2023{\natexlab{c}}, arXiv e-prints,
  arXiv:2309.01325

\bibitem[{{Guo} {et~al.}(2016){Guo}, {Xia}, {Keppens}, \& {Valori}}]{Guo2016b}
{Guo}, Y., {Xia}, C., {Keppens}, R., \& {Valori}, G. 2016, \apj, 828, 82

\bibitem[{{Guo} {et~al.}(2019){Guo}, {Xu}, {Ding}, {Chen}, {Xia}, \&
  {Keppens}}]{Guoy2019}
{Guo}, Y., {Xu}, Y., {Ding}, M.~D., {et~al.} 2019, \apjl, 884, L1

\bibitem[{{Guo} {et~al.}(2021{\natexlab{b}}){Guo}, {Zhong}, {Ding}, {Chen},
  {Xia}, \& {Keppens}}]{Guoy2021}
{Guo}, Y., {Zhong}, Z., {Ding}, M.~D., {et~al.} 2021{\natexlab{b}}, \apj, 919,
  39

\bibitem[{{Hirayama}(1974)}]{Hirayama1974}
{Hirayama}, T. 1974, \solphys, 34, 323

\bibitem[{{Hood} \& {Priest}(1981)}]{Hood&Priest1981}
{Hood}, A.~W. \& {Priest}, E.~R. 1981, Geophysical and Astrophysical Fluid
  Dynamics, 17, 297

\bibitem[{{Illing} \& {Hundhausen}(1986)}]{Illing1986}
{Illing}, R.~M.~E. \& {Hundhausen}, A.~J. 1986, \jgr, 91, 10951

\bibitem[{{Isavnin}(2016)}]{Isavnin2016}
{Isavnin}, A. 2016, \apj, 833, 267

\bibitem[{{Jin} {et~al.}(2012){Jin}, {Manchester}, {van der Holst},
  {Gruesbeck}, {Frazin}, {Landi}, {Vasquez}, {Lamy}, {Llebaria}, {Fedorov},
  {Toth}, \& {Gombosi}}]{Jin2012}
{Jin}, M., {Manchester}, W.~B., {van der Holst}, B., {et~al.} 2012, \apj, 745,
  6

\bibitem[{{Jin} {et~al.}(2013){Jin}, {Manchester}, {van der Holst}, {Oran},
  {Sokolov}, {Toth}, {Liu}, {Sun}, \& {Gombosi}}]{Jin2013}
{Jin}, M., {Manchester}, W.~B., {van der Holst}, B., {et~al.} 2013, \apj, 773,
  50

\bibitem[{{Jin} {et~al.}(2017){Jin}, {Manchester}, {van der Holst}, {Sokolov},
  {T{\'o}th}, {Vourlidas}, {de Koning}, \& {Gombosi}}]{Jin2017}
{Jin}, M., {Manchester}, W.~B., {van der Holst}, B., {et~al.} 2017, \apj, 834,
  172

\bibitem[{{Kataoka} {et~al.}(2009){Kataoka}, {Ebisuzaki}, {Kusano}, {Shiota},
  {Inoue}, {Yamamoto}, \& {Tokumaru}}]{Kataoka2006}
{Kataoka}, R., {Ebisuzaki}, T., {Kusano}, K., {et~al.} 2009, Journal of
  Geophysical Research (Space Physics), 114, A10102

\bibitem[{{Kilpua} {et~al.}(2017){Kilpua}, {Koskinen}, \&
  {Pulkkinen}}]{Kilpua2017}
{Kilpua}, E., {Koskinen}, H. E.~J., \& {Pulkkinen}, T.~I. 2017, Living Reviews
  in Solar Physics, 14, 5

\bibitem[{Kimpe {et~al.}(2005)Kimpe, Lani, Quintino, Poedts, \&
  Vandewalle}]{Kimpe2005}
Kimpe, D., Lani, A., Quintino, T., Poedts, S., \& Vandewalle, S. 2005, in
  Recent Advances in Parallel Virtual Machine and Message Passing Interface,
  ed. B.~Di~Martino, D.~Kranzlm{\"u}ller, \& J.~Dongarra (Berlin, Heidelberg:
  Springer Berlin Heidelberg), 520--527

\bibitem[{{Kliem} \& {T{\"o}r{\"o}k}(2006)}]{Kliem2006}
{Kliem}, B. \& {T{\"o}r{\"o}k}, T. 2006, \prl, 96, 255002

\bibitem[{{Kopp} \& {Pneuman}(1976)}]{Kopp1976}
{Kopp}, R.~A. \& {Pneuman}, G.~W. 1976, \solphys, 50, 85

\bibitem[{{Korreck} {et~al.}(2020){Korreck}, {Szabo}, {Nieves Chinchilla},
  {Lavraud}, {Luhmann}, {Niembro}, {Higginson}, {Alzate}, {Wallace}, {Paulson},
  {Rouillard}, {Kouloumvakos}, {Poirier}, {Kasper}, {Case}, {Stevens}, {Bale},
  {Pulupa}, {Whittlesey}, {Livi}, {Goetz}, {Larson}, {Malaspina}, {Morgan},
  {Narock}, {Schwadron}, {Bonnell}, {Harvey}, \& {Wygant}}]{Korreck2020}
{Korreck}, K.~E., {Szabo}, A., {Nieves Chinchilla}, T., {et~al.} 2020, \apjs,
  246, 69

\bibitem[{{Koumtzis} \& {Wiegelmann}(2023)}]{Koumtzis2023}
{Koumtzis}, A. \& {Wiegelmann}, T. 2023, \solphys, 298, 20

\bibitem[{{Krucker} {et~al.}(2011){Krucker}, {Kontar}, {Christe}, {Glesener},
  \& {Lin}}]{Krucker2011}
{Krucker}, S., {Kontar}, E.~P., {Christe}, S., {Glesener}, L., \& {Lin}, R.~P.
  2011, \apj, 742, 82

\bibitem[{{Ku{\'z}ma} {et~al.}(2023){Ku{\'z}ma}, {Brchnelova}, {Perri},
  {Baratashvili}, {Zhang}, {Lani}, \& {Poedts}}]{Kuzma2023}
{Ku{\'z}ma}, B., {Brchnelova}, M., {Perri}, B., {et~al.} 2023, \apj, 942, 31

\bibitem[{Lani {et~al.}(2005)Lani, Quintino, Kimpe, Deconinck, Vandewalle, \&
  Poedts}]{Lani2005}
Lani, A., Quintino, T., Kimpe, D., {et~al.} 2005, in Computational Science --
  ICCS 2005, ed. V.~S. Sunderam, G.~D. van Albada, P.~M.~A. Sloot, \& J.~J.
  Dongarra (Berlin, Heidelberg: Springer Berlin Heidelberg), 279--286

\bibitem[{Lani {et~al.}(2013)Lani, Villedieu, Bensassi, Kapa, Vymazal, Yalim,
  \& Panesi}]{Lani2013}
Lani, A., Villedieu, N., Bensassi, K., {et~al.} 2013, in AIAA 2013-2589, 21th
  AIAA CFD Conference, San Diego (CA)

\bibitem[{{Lani} {et~al.}(2014){Lani}, {Yalim}, \& {Poedts}}]{Lani2014}
{Lani}, A., {Yalim}, M.~S., \& {Poedts}, S. 2014, Computer Physics
  Communications, 185, 2538

\bibitem[{{Lin} {et~al.}(2005){Lin}, {Ko}, {Sui}, {Raymond}, {Stenborg},
  {Jiang}, {Zhao}, \& {Mancuso}}]{Lin2005}
{Lin}, J., {Ko}, Y.~K., {Sui}, L., {et~al.} 2005, \apj, 622, 1251

\bibitem[{{Linan} {et~al.}(2023){Linan}, {Regnault}, {Perri}, {Brchnelova},
  {Kuzma}, {Lani}, {Poedts}, \& {Schmieder}}]{Linan2023}
{Linan}, L., {Regnault}, F., {Perri}, B., {et~al.} 2023, \aap, 675, A101

\bibitem[{{Linton} \& {Antiochos}(2002)}]{Linton2002}
{Linton}, M.~G. \& {Antiochos}, S.~K. 2002, \apj, 581, 703

\bibitem[{{Liu}(2020)}]{Liur2020}
{Liu}, R. 2020, Research in Astronomy and Astrophysics, 20, 165

\bibitem[{{Liu} {et~al.}(2018){Liu}, {Liu}, {Hu}, {Wang}, \& {Zhao}}]{Liuy2018}
{Liu}, Y.~A., {Liu}, Y.~D., {Hu}, H., {Wang}, R., \& {Zhao}, X. 2018, \apj,
  854, 126

\bibitem[{{Lugaz} {et~al.}(2011){Lugaz}, {Downs}, {Shibata}, {Roussev}, {Asai},
  \& {Gombosi}}]{Lugaz2011}
{Lugaz}, N., {Downs}, C., {Shibata}, K., {et~al.} 2011, \apj, 738, 127

\bibitem[{{Lynch} {et~al.}(2009){Lynch}, {Antiochos}, {Li}, {Luhmann}, \&
  {DeVore}}]{Lynch2009}
{Lynch}, B.~J., {Antiochos}, S.~K., {Li}, Y., {Luhmann}, J.~G., \& {DeVore},
  C.~R. 2009, \apj, 697, 1918

\bibitem[{{Maharana} {et~al.}(2022){Maharana}, {Isavnin}, {Scolini}, {Wijsen},
  {Rodriguez}, {Mierla}, {Magdaleni{\'c}}, \& {Poedts}}]{Maharana2021}
{Maharana}, A., {Isavnin}, A., {Scolini}, C., {et~al.} 2022, Advances in Space
  Research, 70, 1641

\bibitem[{{Maharana} {et~al.}(2023){Maharana}, {Scolini}, {Schmieder}, \&
  {Poedts}}]{Maharana2023}
{Maharana}, A., {Scolini}, C., {Schmieder}, B., \& {Poedts}, S. 2023, arXiv
  e-prints, arXiv:2305.06881

\bibitem[{{Masson} {et~al.}(2013){Masson}, {Antiochos}, \&
  {DeVore}}]{Masson2013}
{Masson}, S., {Antiochos}, S.~K., \& {DeVore}, C.~R. 2013, \apj, 771, 82

\bibitem[{{Miki{\'c}} {et~al.}(2018){Miki{\'c}}, {Downs}, {Linker}, {Caplan},
  {Mackay}, {Upton}, {Riley}, {Lionello}, {T{\"o}r{\"o}k}, {Titov}, {Wijaya},
  {Druckm{\"u}ller}, {Pasachoff}, \& {Carlos}}]{Mikic2018}
{Miki{\'c}}, Z., {Downs}, C., {Linker}, J.~A., {et~al.} 2018, Nature Astronomy,
  2, 913

\bibitem[{{Miki{\'c}} {et~al.}(1999){Miki{\'c}}, {Linker}, {Schnack},
  {Lionello}, \& {Tarditi}}]{Mikic1999}
{Miki{\'c}}, Z., {Linker}, J.~A., {Schnack}, D.~D., {Lionello}, R., \&
  {Tarditi}, A. 1999, Physics of Plasmas, 6, 2217

\bibitem[{{Morgan}(2015)}]{Morgan2015}
{Morgan}, H. 2015, \apjs, 219, 23

\bibitem[{{Nakamura} {et~al.}(2011){Nakamura}, {Imamura}, {Ishii}, {Abe},
  {Satoh}, {Suzuki}, {Ueno}, {Yamazaki}, {Iwagami}, {Watanabe}, {Taguchi},
  {Fukuhara}, {Takahashi}, {Yamada}, {Hoshino}, {Ohtsuki}, {Uemizu},
  {Hashimoto}, {Takagi}, {Matsuda}, {Ogohara}, {Sato}, {Kasaba}, {Kouyama},
  {Hirata}, {Nakamura}, {Yamamoto}, {Okada}, {Horinouchi}, {Yamamoto}, \&
  {Hayashi}}]{Nakamura2011}
{Nakamura}, M., {Imamura}, T., {Ishii}, N., {et~al.} 2011, Earth, Planets and
  Space, 63, 443

\bibitem[{{Ni} {et~al.}(2012){Ni}, {Ziegler}, {Huang}, {Lin}, \&
  {Mei}}]{Ni2012}
{Ni}, L., {Ziegler}, U., {Huang}, Y.-M., {Lin}, J., \& {Mei}, Z. 2012, Physics
  of Plasmas, 19, 072902

\bibitem[{{Odstrcil}(2003)}]{Odstrcil2003}
{Odstrcil}, D. 2003, Advances in Space Research, 32, 497

\bibitem[{{Ouyang} {et~al.}(2017){Ouyang}, {Zhou}, {Chen}, \&
  {Fang}}]{Ouyang2017}
{Ouyang}, Y., {Zhou}, Y.~H., {Chen}, P.~F., \& {Fang}, C. 2017, \apj, 835, 94

\bibitem[{{Perri} {et~al.}(2023){Perri}, {Ku{\'z}ma}, {Brchnelova},
  {Baratashvili}, {Zhang}, {Leitner}, {Lani}, \& {Poedts}}]{Perri2023}
{Perri}, B., {Ku{\'z}ma}, B., {Brchnelova}, M., {et~al.} 2023, \apj, 943, 124

\bibitem[{{Perri} {et~al.}(2022){Perri}, {Leitner}, {Brchnelova},
  {Baratashvili}, {Ku{\'z}ma}, {Zhang}, {Lani}, \& {Poedts}}]{PerriLeitner2022}
{Perri}, B., {Leitner}, P., {Brchnelova}, M., {et~al.} 2022, \apj, 936, 19

\bibitem[{{Pesnell} {et~al.}(2012){Pesnell}, {Thompson}, \&
  {Chamberlin}}]{Pesnell2012}
{Pesnell}, W.~D., {Thompson}, B.~J., \& {Chamberlin}, P.~C. 2012, \solphys,
  275, 3

\bibitem[{{Poedts} {et~al.}(2020){Poedts}, {Lani}, {Scolini}, {Verbeke},
  {Wijsen}, {Lapenta}, {Laperre}, {Millas}, {Innocenti}, {Chan{\'e}},
  {Baratashvili}, {Samara}, {Van der Linden}, {Rodriguez}, {Vanlommel},
  {Vainio}, {Afanasiev}, {Kilpua}, {Pomoell}, {Sarkar}, {Aran}, {Sanahuja},
  {Paredes}, {Clarke}, {Thomson}, {Rouilard}, {Pinto}, {Marchaudon}, {Blelly},
  {Gorce}, {Plotnikov}, {Kouloumvakos}, {Heber}, {Herbst}, {Kochanov},
  {Raeder}, \& {Depauw}}]{Poedts2020}
{Poedts}, S., {Lani}, A., {Scolini}, C., {et~al.} 2020, Journal of Space
  Weather and Space Climate, 10, 57

\bibitem[{{Pomoell} \& {Poedts}(2018)}]{Pomoell2019}
{Pomoell}, J. \& {Poedts}, S. 2018, Journal of Space Weather and Space Climate,
  8, A35

\bibitem[{{Regnault} {et~al.}(2020){Regnault}, {Janvier}, {D{\'e}moulin},
  {Auch{\`e}re}, {Strugarek}, {Dasso}, \& {No{\^u}s}}]{Regnault2020}
{Regnault}, F., {Janvier}, M., {D{\'e}moulin}, P., {et~al.} 2020, Journal of
  Geophysical Research (Space Physics), 125, e28150

\bibitem[{{Regnault} {et~al.}(2023){Regnault}, {Strugarek}, {Janvier},
  {Auch{\`e}re}, {Lugaz}, \& {Al-Haddad}}]{Regnault2023}
{Regnault}, F., {Strugarek}, A., {Janvier}, M., {et~al.} 2023, \aap, 670, A14

\bibitem[{{Scherrer} {et~al.}(2012){Scherrer}, {Schou}, {Bush}, {Kosovichev},
  {Bogart}, {Hoeksema}, {Liu}, {Duvall}, {Zhao}, {Title}, {Schrijver},
  {Tarbell}, \& {Tomczyk}}]{Scherrer2012}
{Scherrer}, P.~H., {Schou}, J., {Bush}, R.~I., {et~al.} 2012, \solphys, 275,
  207

\bibitem[{{Schmieder} {et~al.}(2015){Schmieder}, {Aulanier}, \&
  {Vr{\v{s}}nak}}]{Schmieder2015}
{Schmieder}, B., {Aulanier}, G., \& {Vr{\v{s}}nak}, B. 2015, \solphys, 290,
  3457

\bibitem[{{Schrijver} {et~al.}(2015){Schrijver}, {Kauristie}, {Aylward},
  {Denardini}, {Gibson}, {Glover}, {Gopalswamy}, {Grande}, {Hapgood},
  {Heynderickx}, {Jakowski}, {Kalegaev}, {Lapenta}, {Linker}, {Liu},
  {Mandrini}, {Mann}, {Nagatsuma}, {Nandy}, {Obara}, {Paul O'Brien}, {Onsager},
  {Opgenoorth}, {Terkildsen}, {Valladares}, \& {Vilmer}}]{Schrijver2015}
{Schrijver}, C.~J., {Kauristie}, K., {Aylward}, A.~D., {et~al.} 2015, Advances
  in Space Research, 55, 2745

\bibitem[{{Scolini} {et~al.}(2020){Scolini}, {Chan{\'e}}, {Temmer}, {Kilpua},
  {Dissauer}, {Veronig}, {Palmerio}, {Pomoell}, {Dumbovi{\'c}}, {Guo},
  {Rodriguez}, \& {Poedts}}]{Scolini2020}
{Scolini}, C., {Chan{\'e}}, E., {Temmer}, M., {et~al.} 2020, \apjs, 247, 21

\bibitem[{{Scolini} {et~al.}(2019){Scolini}, {Rodriguez}, {Mierla}, {Pomoell},
  \& {Poedts}}]{Scolini2019}
{Scolini}, C., {Rodriguez}, L., {Mierla}, M., {Pomoell}, J., \& {Poedts}, S.
  2019, \aap, 626, A122

\bibitem[{{Shen} {et~al.}(2011){Shen}, {Wang}, {Gui}, {Ye}, \&
  {Wang}}]{Shen2011}
{Shen}, C., {Wang}, Y., {Gui}, B., {Ye}, P., \& {Wang}, S. 2011, \solphys, 269,
  389

\bibitem[{{Shen} {et~al.}(2022){Shen}, {Shen}, {Xu}, {Liu}, {Feng}, \&
  {Wang}}]{Shen2022}
{Shen}, F., {Shen}, C., {Xu}, M., {et~al.} 2022, Reviews of Modern Plasma
  Physics, 6, 8

\bibitem[{{Shen} {et~al.}(2014){Shen}, {Shen}, {Zhang}, {Hess}, {Wang}, {Feng},
  {Cheng}, \& {Yang}}]{Shen2014}
{Shen}, F., {Shen}, C., {Zhang}, J., {et~al.} 2014, Journal of Geophysical
  Research (Space Physics), 119, 7128

\bibitem[{{Shiota} {et~al.}(2005){Shiota}, {Isobe}, {Chen}, {Yamamoto},
  {Sakajiri}, \& {Shibata}}]{Shiota2005}
{Shiota}, D., {Isobe}, H., {Chen}, P.~F., {et~al.} 2005, \apj, 634, 663

\bibitem[{{Shiota} \& {Kataoka}(2016)}]{Shiota2016}
{Shiota}, D. \& {Kataoka}, R. 2016, Space Weather, 14, 56

\bibitem[{{Shiota} {et~al.}(2010){Shiota}, {Kusano}, {Miyoshi}, \&
  {Shibata}}]{Shiota2010}
{Shiota}, D., {Kusano}, K., {Miyoshi}, T., \& {Shibata}, K. 2010, \apj, 718,
  1305

\bibitem[{{Singh} {et~al.}(2018){Singh}, {Yalim}, \& {Pogorelov}}]{Singh2018}
{Singh}, T., {Yalim}, M.~S., \& {Pogorelov}, N.~V. 2018, \apj, 864, 18

\bibitem[{{Song} {et~al.}(2022){Song}, {Li}, \& {Chen}}]{Song2022}
{Song}, H., {Li}, L., \& {Chen}, Y. 2022, \apj, 933, 68

\bibitem[{{Sturrock}(1966)}]{Sturrock1966}
{Sturrock}, P.~A. 1966, \nat, 211, 695

\bibitem[{{Titov} \& {D{\'e}moulin}(1999)}]{Titov1999}
{Titov}, V.~S. \& {D{\'e}moulin}, P. 1999, \aap, 351, 707

\bibitem[{{Titov} {et~al.}(2018){Titov}, {Downs}, {Miki{\'c}}, {T{\"o}r{\"o}k},
  {Linker}, \& {Caplan}}]{Titov2018}
{Titov}, V.~S., {Downs}, C., {Miki{\'c}}, Z., {et~al.} 2018, \apjl, 852, L21

\bibitem[{{Titov} {et~al.}(2014){Titov}, {T{\"o}r{\"o}k}, {Mikic}, \&
  {Linker}}]{Titov2014}
{Titov}, V.~S., {T{\"o}r{\"o}k}, T., {Mikic}, Z., \& {Linker}, J.~A. 2014,
  \apj, 790, 163

\bibitem[{{T{\"o}r{\"o}k} {et~al.}(2010){T{\"o}r{\"o}k}, {Berger}, \&
  {Kliem}}]{Torok2010}
{T{\"o}r{\"o}k}, T., {Berger}, M.~A., \& {Kliem}, B. 2010, \aap, 516, A49

\bibitem[{{T{\"o}r{\"o}k} {et~al.}(2018){T{\"o}r{\"o}k}, {Downs}, {Linker},
  {Lionello}, {Titov}, {Miki{\'c}}, {Riley}, {Caplan}, \& {Wijaya}}]{Torok2018}
{T{\"o}r{\"o}k}, T., {Downs}, C., {Linker}, J.~A., {et~al.} 2018, \apj, 856, 75

\bibitem[{{T{\"o}r{\"o}k} {et~al.}(2004){T{\"o}r{\"o}k}, {Kliem}, \&
  {Titov}}]{Torok2004}
{T{\"o}r{\"o}k}, T., {Kliem}, B., \& {Titov}, V.~S. 2004, \aap, 413, L27

\bibitem[{{T{\'o}th} {et~al.}(2012){T{\'o}th}, {van der Holst}, {Sokolov}, {De
  Zeeuw}, {Gombosi}, {Fang}, {Manchester}, {Meng}, {Najib}, {Powell}, {Stout},
  {Glocer}, {Ma}, \& {Opher}}]{Toth2012}
{T{\'o}th}, G., {van der Holst}, B., {Sokolov}, I.~V., {et~al.} 2012, Journal
  of Computational Physics, 231, 870

\bibitem[{{Tsurutani} {et~al.}(2023){Tsurutani}, {Zank}, {Sterken}, {Shibata},
  {Nagai}, {Mannucci}, {Malaspina}, {Lakhina}, {Kanekal}, {Hosokawa}, {Horne},
  {Hajra}, {Glassmeier}, {Gaunt}, {Chen}, \& {Akasofu}}]{Tsurutani2023}
{Tsurutani}, B.~T., {Zank}, G.~P., {Sterken}, V.~J., {et~al.} 2023, IEEE
  Transactions on Plasma Science, 51, 1595

\bibitem[{{van der Holst} {et~al.}(2010){van der Holst}, {Manchester},
  {Frazin}, {V{\'a}squez}, {T{\'o}th}, \& {Gombosi}}]{VanderHolst2010}
{van der Holst}, B., {Manchester}, W.~B., I., {Frazin}, R.~A., {et~al.} 2010,
  \apj, 725, 1373

\bibitem[{{Vandenhoeck} \& {Lani}(2019)}]{Vandenhoeck2019}
{Vandenhoeck}, R. \& {Lani}, A. 2019, Computer Physics Communications, 242, 1

\bibitem[{{Verbeke} {et~al.}(2022){Verbeke}, {Baratashvili}, \&
  {Poedts}}]{Verbeke2022}
{Verbeke}, C., {Baratashvili}, T., \& {Poedts}, S. 2022, \aap, 662, A50

\bibitem[{{Verbeke} {et~al.}(2019){Verbeke}, {Pomoell}, \&
  {Poedts}}]{Verbeke2019}
{Verbeke}, C., {Pomoell}, J., \& {Poedts}, S. 2019, \aap, 627, A111

\bibitem[{{Vourlidas} \& {Howard}(2006)}]{Vourlidas2006}
{Vourlidas}, A. \& {Howard}, R.~A. 2006, \apj, 642, 1216

\bibitem[{{Vourlidas} {et~al.}(2013){Vourlidas}, {Lynch}, {Howard}, \&
  {Li}}]{Vourlidas2013}
{Vourlidas}, A., {Lynch}, B.~J., {Howard}, R.~A., \& {Li}, Y. 2013, \solphys,
  284, 179

\bibitem[{{Wang} {et~al.}(2017){Wang}, {Liu}, {Wang}, {Hu}, {Shen}, {Jiang}, \&
  {Zhu}}]{Wangws2017}
{Wang}, W., {Liu}, R., {Wang}, Y., {et~al.} 2017, Nature Communications, 8,
  1330

\bibitem[{{Wang} {et~al.}(2021){Wang}, {Cheng}, {Ding}, \& {Lu}}]{Wang2021}
{Wang}, Y., {Cheng}, X., {Ding}, M., \& {Lu}, Q. 2021, \apj, 923, 227

\bibitem[{{Webb} \& {Howard}(2012)}]{Webb2012}
{Webb}, D.~F. \& {Howard}, T.~A. 2012, Living Reviews in Solar Physics, 9, 3

\bibitem[{{Wiegelmann} \& {Sakurai}(2021)}]{Wiegelmann2021}
{Wiegelmann}, T. \& {Sakurai}, T. 2021, Living Reviews in Solar Physics, 18, 1

\bibitem[{{Xu} {et~al.}(2020){Xu}, {Zhu}, \& {Guo}}]{Xu2020}
{Xu}, Y., {Zhu}, J., \& {Guo}, Y. 2020, \apj, 892, 54

\bibitem[{{Yeates} \& {Hornig}(2016)}]{Yeats2016}
{Yeates}, A.~R. \& {Hornig}, G. 2016, \aap, 594, A98

\bibitem[{{Zhang} {et~al.}(2012){Zhang}, {Cheng}, \& {Ding}}]{Zhang2012}
{Zhang}, J., {Cheng}, X., \& {Ding}, M.-D. 2012, Nature Communications, 3, 747

\bibitem[{{Zhou} \& {Feng}(2017)}]{Zhou2017}
{Zhou}, Y. \& {Feng}, X. 2017, Journal of Geophysical Research (Space Physics),
  122, 1451

\bibitem[{{Zhou} {et~al.}(2012){Zhou}, {Feng}, {Wu}, {Du}, {Shen}, \&
  {Xiang}}]{Zhou2012}
{Zhou}, Y.~F., {Feng}, X.~S., {Wu}, S.~T., {et~al.} 2012, Journal of
  Geophysical Research (Space Physics), 117, A01102

\end{thebibliography}

\end{document}